\documentclass{article}

\usepackage{amsmath,amssymb,graphicx,epstopdf}
\usepackage{lineno,natbib}
\usepackage{subfigure,authblk}
\usepackage[margin=1in]{geometry}
\DeclareGraphicsExtensions{.eps}

\def\begineq{\begin{equation}}
\def\endeq{\end{equation}}

\def\beginseq{\begin{subequations}}
\def\endseq{\end{subequations}}

\def\beging{\begin{gather}}
\def\endg{\end{gather}}

\def\begineqn{\begin{equation*}}
\def\endeqn{\end{equation*}}

\def\beginar{\begin{eqnarray}}
\def\endar{\end{eqnarray}}
\def\beginarn{\begin{eqnarray*}}
\def\endarn{\end{eqnarray*}}

\def\lb{\left ( }
\def\rb{\right ) }
\def\lsq{\left [ }
\def\rsq{\right ]}

\def\ep{\epsilon}

\def\ub{\mathbf{u}}

\def\ubp{\mathbf{u}^{\prime}}

\def\rot{\boldsymbol{\widehat{\eta}}}

\def\pfu{u^{\prime}}
\def\pfv{v^{\prime}}
\def\pfw{w^{\prime}}

\def\dsx{{\partial_x}}
\def\dsy{{\partial_y}}

\def\dst{{\partial_t}}

\def\dsz{{\partial_z}}

\def\dszz{{\partial_{zz}}}

\def\hz{{\bf\widehat z}}

\def\mp{\overline{p}}
\def\mrho{\overline{\rho}} 
\def\mT{\overline{T}}
\def\pp{p^{\prime}}
\def\prho{\rho^{\prime}}
\def\pT{T^{\prime}}
\def\pS{S^{\prime}}
\def\mS{\overline{S}}

\def\div{{\nabla \cdot}}
\def\divp{{\nabla_\perp \cdot \,}}

\begin{document}

\title{The Breakdown of the Anelastic Approximation in Rotating Compressible Convection: Implications for Astrophysical Systems}

\author{Michael A. Calkins\thanks{michael.calkins@colorado.edu}}
\author{Keith Julien\thanks{keith.julien@colorado.edu}}
\author{Philippe Marti\thanks{philippe.marti@colorado.edu}}
\affil{Department of Applied Mathematics, University of Colorado, Boulder, CO 80309, USA}
\maketitle

\begin{abstract}
  The linear theory for rotating compressible convection in a plane layer geometry is presented for the astrophysically-relevant case of low Prandtl number gases.  When the rotation rate of the system is large, the flow remains geostrophically balanced for all stratification levels investigated and the classical (i.e., incompressible) asymptotic scaling laws for the critical parameters are recovered.  For sufficiently small Prandtl numbers, increasing stratification tends to further destabilise the fluid layer, decrease the critical wavenumber and increase the oscillation frequency of the convective instability.  In combination, these effects increase the relative magnitude of the time derivative of the density perturbation contained in the conservation of mass equation to non-negligible levels; the resulting convective instabilities occur in the form of compressional quasi-geostrophic oscillations.  We find that the anelastic equations, which neglect this term, cannot capture these instabilities and possess spuriously-growing eigenmodes in the rapidly rotating, low Prandtl number regime.  It is shown that the Mach number for rapidly rotating compressible convection is intrinsically small for all background states, regardless of the departure from adiabaticity.
\end{abstract}

\section{Introduction}
\label{S:intro}

From millimetre-scale acoustic waves to planetary-scale atmospheric jets, the Navier-Stokes equations (NSE) are capable of accurately modelling fluid flows characterised by a broad range of spatial and temporal scales.  This generality, however, also leads to prohibitively high computational costs when solving the governing equations numerically, and prevents direct numerical simulations from accessing flow regimes that are relevant for understanding geophysical and astrophysical fluid systems.  This implicit cost of the general, compressible NSE is the primary motivation for developing simplified, or reduced, forms of the NSE.  

For the particular problem of buoyancy-driven flows, the Oberbeck-Boussinesq equations (OBE) \citep{aO79,jB03} are one of the most well studied reduced forms of the NSE. The main characteristic of the OBE is that the fluid density is treated as a constant value except where it occurs in the buoyancy force; the resulting fluid is incompressible in the sense that acoustic waves are filtered from the dynamics.  The main benefits of reduced equations are that they can often be solved by simpler and faster numerical algorithms, they provide a simplified physical picture of the flows of interest by eliminating dynamically unimportant phenomena (e.g.~acoustic waves), and can thus often lead to a greater amount of insight than what can be gained by analysis of the full NSE.  Because reduced equations are approximations of the NSE, however, they are accurate only for certain ranges of the parameter values relevant to a given problem.  The practical limitations of the OBE are well known, namely that variations of the state variables (i.e., density, pressure and temperature) over the vertical scale of the system must remain small for the OBE to be an accurate approximation of the NSE \citep{eS60,jM62}.  Unfortunately, this requirement is rarely satisfied in natural systems \citep[e.g.][]{jC96,tG05}.

To overcome the practical limitations of the OBE, \cite{gB53} and \cite{yO62} derived the so-called anelastic equations (AE) that allow for vertically stratified state variables, but, like the OBE, filter acoustic waves from the dynamics by eliminating the temporal derivative of the density perturbation present in the conservation of mass equation. The AE are derived by representing each flow variable by an asymptotic expansion, with the small parameter representing the deviation from an adiabatic background state.  In non-dimensional terms, the small parameter is the (squared) Mach number, a ratio of a characteristic flow speed to the sound speed \citep[see][]{dG69,prB96,rK10}.  An important component of the AE is that the convective fluctuations do not feed back onto the background state.  

The AE are now a commonly employed tool for theoretical and numerical investigations of both stably stratified gravity wave dynamics \citep{tL12,bB12} and unstably stratified convection dynamics \citep{cJ11}.  Until recently, however, few investigations have been carried out that make one-to-one comparisons between the NSE and the AE for the case of compressible convection \citep[e.g.~see][]{nB10,mC14,dL14}.

One of the most common systems in which to study rotating convection is the plane layer geometry, or rotating Rayleigh-B\'enard configuration, in which  a constant temperature difference $\Delta T$ is held across two horizontal boundaries separated by a vertical distance $H$, and the system rotates about the vertical axis with rotation vector $\mathbf{\Omega} = \Omega \hz$.  Here the $z$-axis is measured positively downwards such that for a constant gravity vector $g = g \hz$, the fluid layer becomes convectively unstable for a sufficiently large negative heat flux.  Much of what is currently known about rotating convection has been obtained from studying the OBE within the Rayleigh-B\'enard configuration \citep[e.g.][]{sC61}.  For the OBE, the dynamics of the system are completely specified by three independent dimensionless parameters
\begineq
Ta_I = \lb \frac{2 \Omega H^2}{\nu} \rb^2 = \lb \frac{\textnormal{Coriolis force}}{\textnormal{viscous force}} \rb^2, 
\endeq
\begineq
Ra_I = \frac{g \alpha \Delta T H^3}{\nu \kappa} = \frac{\textnormal{buoyancy}}{\textnormal{diffusion}},
\endeq
\begineq
Pr_{I} = \frac{\nu}{\kappa} = \frac{\textnormal{viscous diffusion}}{\textnormal{thermal diffusion}},
\endeq
where $\nu$ is the kinematic viscosity, $\alpha$ is the thermal expansion coefficient, and $\kappa$ is the thermal diffusivity.  The subscript ``$I$'' used for the above definitions refers to parameters employed for an incompressible fluid.  The Taylor number ($Ta_I$) represents the (squared) ratio of Coriolis force to the viscous force, the Rayleigh number ($Ra_I$) is the strength of the thermal forcing relative to diffusive effects, and the Prandtl number ($Pr_I$) is an intrinsic fluid property that is defined as the ratio of viscous to thermal diffusion.  All geophysical and astrophysical convecting fluids are characterised by $Ta_I \gg 1$ and $Ra_I \gg 1$. Most geophysical and astrophysical gases are characterised by Prandtl numbers less than unity, with recent calculations suggesting that the Prandtl number in Jupiter may be as small as $O(10^{-2})$ \citep{mF12}, and stellar convection zones having Prandtl numbers as small as $10^{-6}$ \citep[e.g][]{mM05}. For comparison, water at standard temperature and pressure is characterised by $Pr_I \approx 7$.

When the Taylor number is large and the Prandtl number is order unity and greater, the most unstable convective instabilities take the form of steady roll-like patterns with a critical wavenumber that scales as $k_c \sim Ta_I^{1/6}$ and a critical Rayleigh number that scales as $Ra_c \sim Ta_I^{2/3}$ \citep{sC61}.  For $Pr_I \lesssim 0.68$, the preferred instability becomes oscillatory with critical parameters that are Prandtl dependent; when $Pr_I \ll 1$ (but $Pr_I \gg Ta_I^{-1/2}$, \citep[e.g.~see][]{kZ97b}), it can be shown that $\omega_c \sim (Ta_I/Pr_I)^{1/3}$, $k_c \sim Pr_I^{1/3} Ta_I^{1/6}$, and $Ra_c \sim Pr_I^{4/3} Ta_I^{2/3}$. Importantly, the asymptotic scalings of the critical parameters for both $Pr \sim O(1)$ and $Pr \ll 1$ fluids arise when the convecting fluid is geostrophically balanced to leading order \citep[e.g.][]{mS06}.  If the Prandtl number is so small that the distinguished limit $Pr_I \sim O(Ta_I^{-1/2})$ is satisfied, the inertial acceleration in the momentum equations becomes of the same order as the Coriolis and pressure gradient forces and thus the convection is no longer balanced; this regime is characterized by thermally-driven inertial waves with critical parameters $k_c \sim O(1)$, $Ra_c \sim Ta^{-1/2}$ and $\omega_c \sim Ta_I^{1/2}$ \citep{sC61,kZ97b}.  

Recently it was shown that the $Pr \sim O(1)$ quasi-geostrophic convection regime also occurs in both compressible and anelastic ideal gases \cite{mC14}.  With the exception of the anelastic study of \cite{sD95}, there have been no investigations of rapidly rotating compressible convection in low Prandtl number gases, despite the astrophysical relevance of this regime.  In the present work we investigate this parameter regime with both the NSE and the AE, and show that the fundamental instability consists of compressional quasi-geostrophic (low frequency) oscillations whose existence intrinsically depends upon the presence of the time derivative of the density perturbation present in the mass conservation equation.  Because of this dependence, the AE are shown to fail in this regime.

\section{Governing Equations}

In the present work, we present \textit{linear} numerical results obtained from both the NSE and the AE; in both equation sets we assume a calorically perfect Newtonian gas.  All fields are decomposed into convective perturbations and a horizontally averaged background state.  For the NSE the perturbations are denoted with a prime $(\cdot )^\prime$ and the background state is denoted with an overbar $\overline{(\cdot)}$.  Written in non-dimensional form, the NSE in a rotating reference frame are 
\begineq 
\begin{split}
\mrho \lb \dst \ubp + \sqrt{\frac{Pr Ta_o}{Ra_o}} \, \hz \times \ubp \rb = -  H_s \nabla \pp + H_s  \prho \, \hz  + \\ \sqrt{\frac{Pr}{Ra_o}} \lsq \nabla^2 \ubp + \frac{1}{3} \nabla \lb \nabla \cdot \ubp \rb \rsq , 
\label{E:compmom1} 
\end{split}
\endeq
\begineq
\dst \prho + \div \lb \mrho \ubp \rb = 0, 
\label{E:mass}
\endeq
\begineq
\mrho \mT \lb \dst \pS + \pfw \dsz \mS \rb = \frac{1}{\sqrt{Pr Ra_o}} \nabla^2 \pT , 
\endeq
\begineq
\mp = H_a \lb \frac{\gamma-1}{\gamma} \rb \mrho \mT , \quad \frac{\pp}{\mp} = \frac{\prho}{\mrho} + \frac{\pT}{\mT},  \label{E:gas} 
\endeq
\begineq
\mS = \ln \lb \frac{\mp^{1/\gamma}}{\mrho} \rb, \quad \pS = \frac{\pp}{\gamma \mp} - \frac{\prho}{\mrho} \label{E:ent1} ,
\endeq
where the density, velocity vector, pressure, temperature and entropy are denoted by $\rho$, $\mathbf{u}$, $p$, $T$ and $S$, respectively.  The equations have been non-dimensionalised with the following dimensional scales \citep[c.f.][]{mC14} 
\begineq
u \sim  \lb \frac{\beta g H^2}{T_{o}} \rb^{1/2}, \quad t \sim \frac{H}{u}, \quad p \sim \rho_o g H, \quad \rho \sim \rho_{o},
\endeq
\begineq
 T \sim T_{o}, \quad S \sim c_p ,
\endeq
where $\beta$ is the superadiabatic temperature gradient across the fluid layer, and $c_p$ is the specific heat at constant pressure.  Quantities with the subscript ``o'' denote values evaluated at the upper (or outer) boundary.  The above velocity scaling is typically referred to as the convective ``free-fall'' velocity and can be obtained from the momentum equations by balancing inertia with buoyancy.  The compressible versions of the Rayleigh, Taylor and Prandtl numbers are defined by
\begineq
Ra_o = \frac{\rho_o^2 c_p g \beta H^4}{T_{o} \mu k}, \quad Ta_o = \lb \frac{2 \rho_o \Omega H^2}{\mu} \rb^2, \quad Pr = \frac{\mu c_p}{k} .
\endeq
We assume the dynamic viscosity, $\mu$, and the thermal conductivity, $k$, are constant. In addition to the Rayleigh, Taylor, and Prandtl numbers, three additional non-dimensional parameters are required to completely specify the problem of rotating compressible convection; these quantities are the superadiabitic temperature scale height, the adiabatic temperature scale height, and the ratio of specific heats, defined by, respectively,
\begineq
H_s = \frac{T_{o}}{\beta H},  \quad H_a = \frac{c_p T_{o}}{g H}, \quad \gamma = \frac{c_p}{c_v}.
\endeq
We note that the NSE can be reduced to the OBE by taking the double asymptotic limit $H_a^{-1} \rightarrow 0$ and $H_s^{-1} \rightarrow 0$.  

Another important dimensionless parameter that is present in equation \eqref{E:compmom1} is the convective Rossby number defined by
\begineq
Ro  = \sqrt{\frac{Ra_o}{Pr Ta_o}}. 
\endeq
It is important to emphasize that $Ro$ is not an independent dimensionless parameter, but serves as a useful measure of the importance of the Coriolis force.  \textit{Rapidly rotating} convection is distinguished by $Ta \gg 1$ and $Ro \ll 1$.  

To investigate the stability of the NSE it is necessary to define the basic state that exists in the absence of convection by solving the governing equations with $\mathbf{u} \equiv 0$, i.e.,
\begineq
\dszz \mT = 0, \quad \dsz \mp = \mrho .
\endeq
Coupled with the mean ideal gas law the above equations yield 
\begineq
\label{E:cback}
\mT = 1 + \lb H_s^{-1} + H_a^{-1} \rb z, \quad
\mp = H_a \lb \frac{\gamma-1}{\gamma} \rb \mT^{n+1} , \quad
\mrho = \mT^n, 
\endeq
where the (non-dimensional) polytropic index is defined by
\begineq
n = \lb \frac{\gamma}{\gamma-1} \rb \frac{H_s}{H_a+H_s} - 1 . \label{E:pindex}
\endeq

In general, the Rayleigh and Taylor numbers are depth-varying quantities and our use of $Ra_o$ and $Ta_o$ in the governing equations is meant for notational convenience only.  All the values reported in the present work are evaluated at the bottom boundary since this is where asymptotic behavior is first observed \citep{wH73}; these values are related by  
\begineq
Ra = \lb 1+ H_s^{-1} + H_a^{-1} \rb^{2n-1} Ra_o, 
\endeq
\begineq
Ta = \lb 1+ H_s^{-1} + H_a^{-1} \rb^{2n} Ta_o.
\endeq

The AE can be obtained directly from the NSE by assuming the flow to be nearly adiabatic in the sense that $\ep = H_s^{-1} \ll 1$ and expanding all dependent variables in an asymptotic series \citep[see][]{yO62}; for instance $\mathbf{u} = \mathbf{u}_0 + \ep \mathbf{u}_1 + \ldots$.  At $O(1)$ we have the background (adiabatic) state defined by
\begineq
\dszz T_0 = 0, \quad \dsz p_0 = \rho_0 .
\endeq
The solutions of which are then
\begineq
T_0 = 1 + H_a^{-1} z, \quad
p_0 = H_a \lb \frac{\gamma-1}{\gamma} \rb T_0^{n_a+1} , \quad
\rho_0 = T_0^{n_a},
\label{E:anelref}
\endeq
where the adiabatic polytropic index is $n_a \equiv (\gamma-1)^{-1}$.  Therefore, thermodynamic variables with the subscript ``0'' represent the adiabatic background state for the AE.  The (linear) AE are obtained at $O(\ep)$ to give
\begineq
\begin{split}
\rho_0 \lb \dst \ub_0 + \sqrt{\frac{Pr Ta^{A}_o}{Ra_o}} \rot \times \ub_0 \rb = -\nabla p_1 + \rho_1 \, \hz  + \\ \sqrt{\frac{Pr}{Ra^A_o}} \lsq \nabla^2 \ub_0 + \frac{1}{3} \nabla \lb \nabla \cdot \ub_0 \rb \rsq , \label{E:anelmom} 
\end{split}
\endeq
\begineq
\div \lb \rho_0 \ub_0 \rb = 0, \label{E:anelcont} 
\endeq
\begineq
\rho_0 T_0 D_t S_1 = \frac{1}{\sqrt{Pr Ra^A_o}} \nabla^2 T_1 , \label{E:enanel} 
\endeq
\begineq
\frac{p_1}{p_0} = \frac{\rho_1}{\rho_0} + \frac{T_1}{T_0}, \label{E:gasanel} 
\endeq
\begineq
S_1 = \frac{p_1}{\gamma p_0} - \frac{\rho_1}{\rho_0} . \label{E:entanel} 
\endeq
For the AE, the Rayleigh and Taylor numbers evaluated at the bottom boundary are given by
\begineq
Ra^A = \lb 1+ H_a^{-1} \rb^{2n_a-1} Ra^A_o, \quad 
\endeq
\begineq
Ta^A = \lb 1+ H_a^{-1} \rb^{2n_a} Ta^A_o.
\endeq

Provided that $n\approx n_a$, the primary difference between the AE and the NSE is the form of the mass conservation equation.  Specifically, note the absence of the $\ep \dst \rho_1$ in equation \eqref{E:anelcont} in comparison to equation \eqref{E:mass}.  Due to the asymptotic reduction necessary to obtain the AE, the number of dimensionless parameters required to specify the anelastic system has been reduced to five ($H_s$ no longer explicitly appears in the governing equations).  

\begin{figure}
  \begin{center}
      \includegraphics[width=7cm]{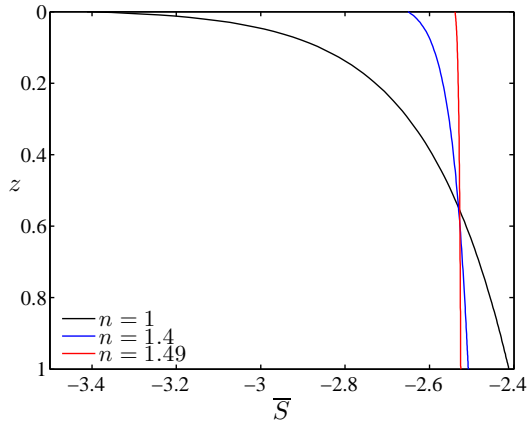}
  \end{center}
  \caption{Vertical profiles of the background entropy $\mS$ used for the NSE for three different values of the polytropic index, $n$.  The present work utilizes a specific heat ratio of $\gamma=5/3$, such that a neutrally stratified, adiabatic background state in which $\mS \rightarrow const.$ is characterized by a polytropic index of $n_a = 1.5$.}
\label{F:entropy}
\end{figure}

Although the equations can be cleanly written in terms of the temperature scale heights $H_s$ and $H_a$, it is perhaps more physically meaningful, and provides a more transparent and broader connection with previous work \citep[e.g.][]{cJ11}, to specify the stratification in terms of the number of density scale heights 
\begineq
N_{\rho} = \log \lb \frac{\rho_i}{\rho_o} \rb,
\endeq
along with the polytropic index.  In all cases presented the specific heat ratio is fixed at $\gamma=5/3$, representing a monotomic ideal gas.  With this specific heat ratio, the adiabatic polytropic index is then $n_a=1.5$.  To drive convection, the fluid layer must be superadiabatic in the sense that $n < n_a$.  We present results from the NSE with three different values of the polytropic index, $n=1$, $n=1.4$, and $n=1.49$.  The case $n=1.49$ results in a background state that differs from an adiabat by $\approx 1 \%$ and thus provides a crucial benchmark for testing the accuracy of the AE.  To illustrate the influence $n$ on the character of the background state, Figure \ref{F:entropy} shows the background entropy $\mS$ for three different values of $n$ where it is observed that $\mS$ approaches a constant value as $n \rightarrow n_a$.

To solve the two equation sets, each flow variable is represented by the typical normal mode ansatz, e.g.$\phantom{a}\mathbf{u} = \widehat{\mathbf{u}}(z) \exp{\lsq i \lb \omega t + \bf{k_\perp} \cdot \bf{x}\rb\rsq} + c.c.$, where $\omega$ is the oscillation frequency and $\mathbf{k_\perp}=(k_x,k_y)$ is the horizontal wavenumber vector.  To obtain spectral accuracy, the equations are discretised in the vertical dimension by an expansion of Cheybshev polynomials.  Up to 80 Chebyshev polynomials were employed to numerically resolve the most extreme cases (i.e., large values of $Ta$ and $N_\rho$).  Constant temperature, stress-free boundary conditions are used for all the reported results.  To generate a numerically sparse system, we use the Chebyshev three-term recurrence relation and solve directly for the spectral coefficients; the boundary conditions are enforced via ``tau"-lines \citep{dG93}.  Due to the background stratification, both the NSE and the AE possess non-constant coefficients terms; these terms are treated efficiently by employing standard convolution operations for the Chebyshev polynomials \citep{gB97,sO13}.  We solve the resulting generalised eigenvalue problem with MATLAB's sparse eigenvalue solver, \textit{sptarn}.  The \textit{critical parameters} are those values associated with the smallest value of the Rayleigh number characterised by a zero growth rate; the resulting parameters are denoted by $Ra_c$, $k_c$ and $\omega_c$.  Further details of the numerical techniques employed can be found in \cite{mC13}.

\section{Results}

\begin{figure}
  \begin{center}
     \subfigure[]{
      \includegraphics[width=7.5cm]{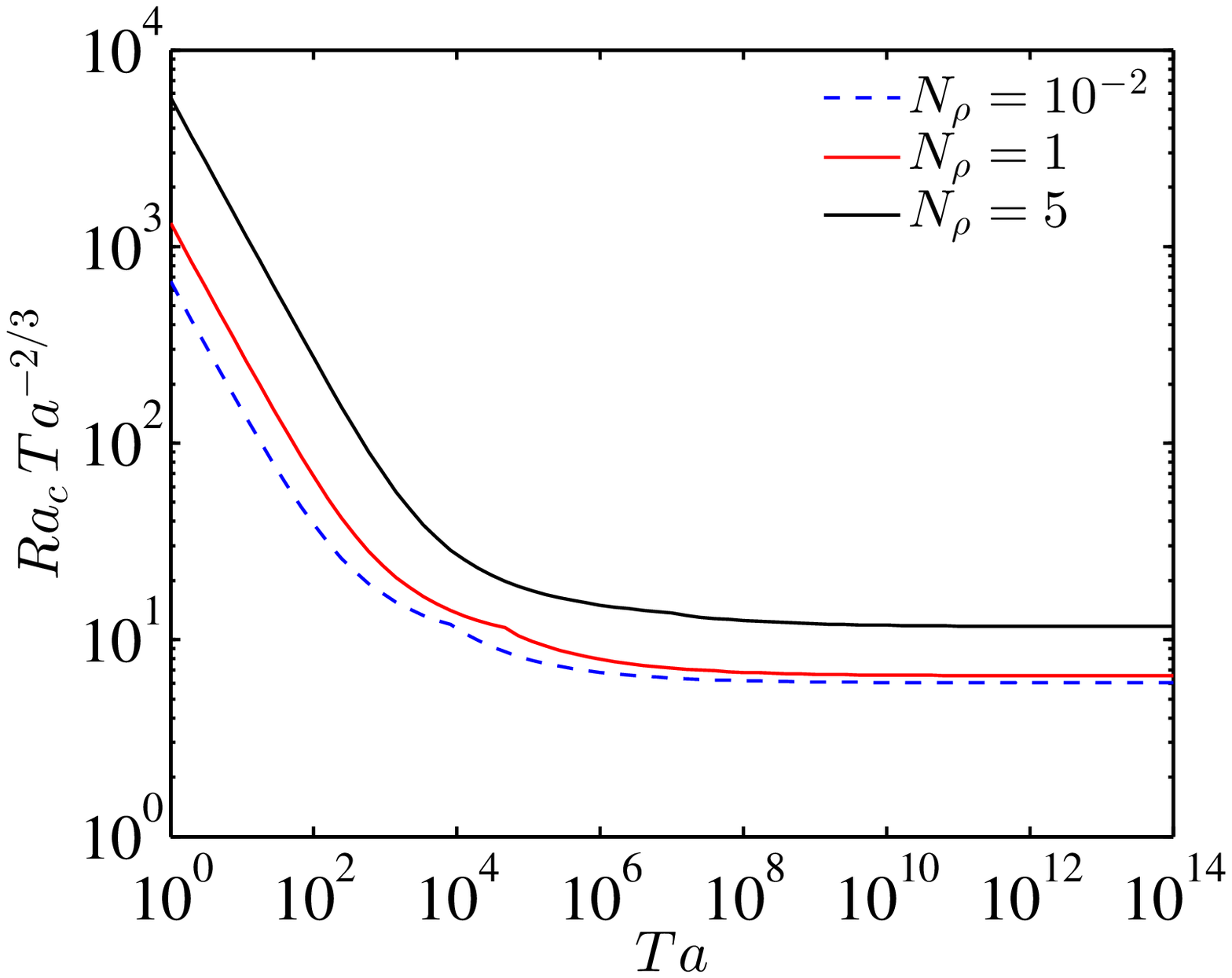}
      } \quad
      \subfigure[]{
      \includegraphics[width=7.5cm]{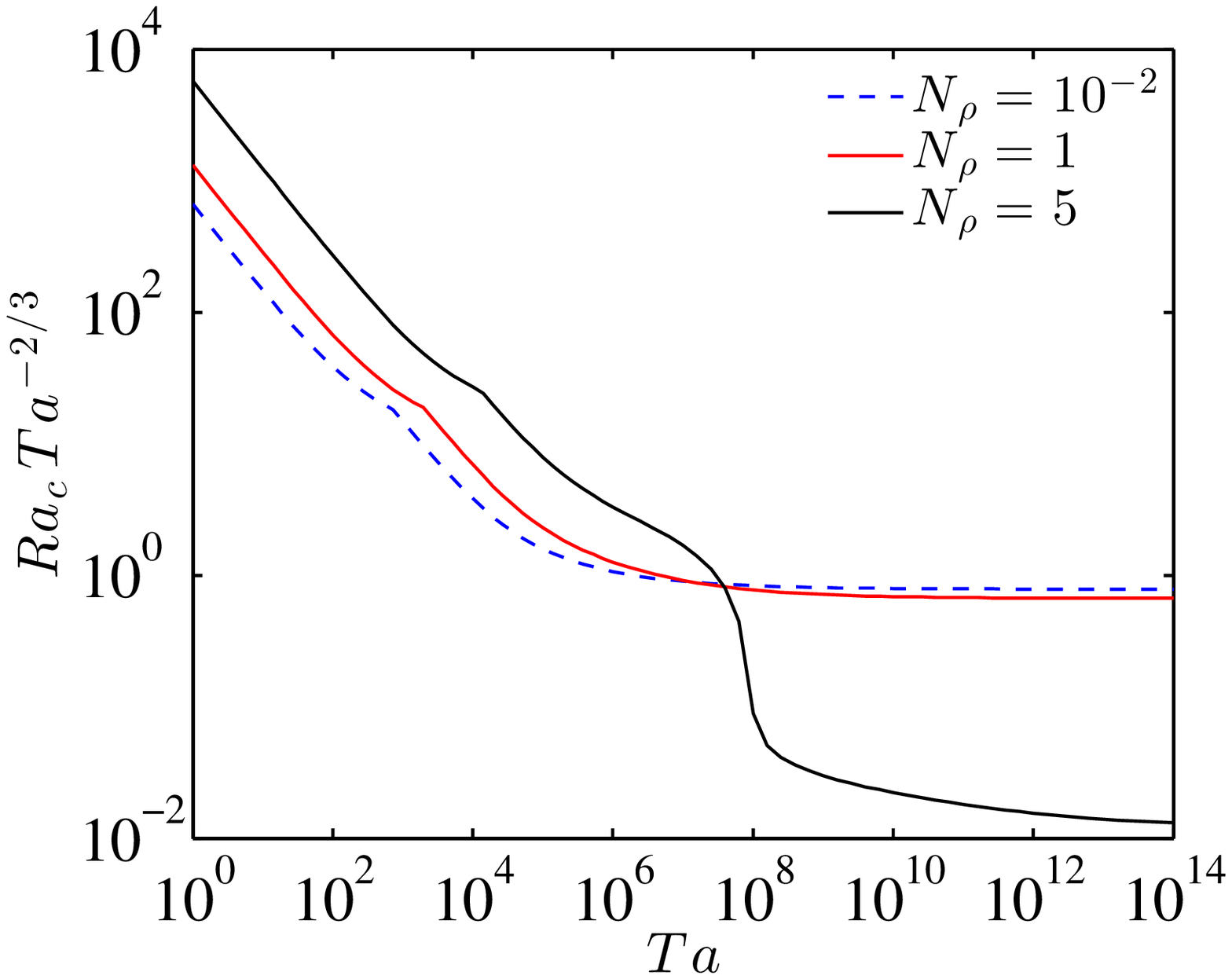}
      } \\   
     \subfigure[]{
      \includegraphics[width=7.5cm]{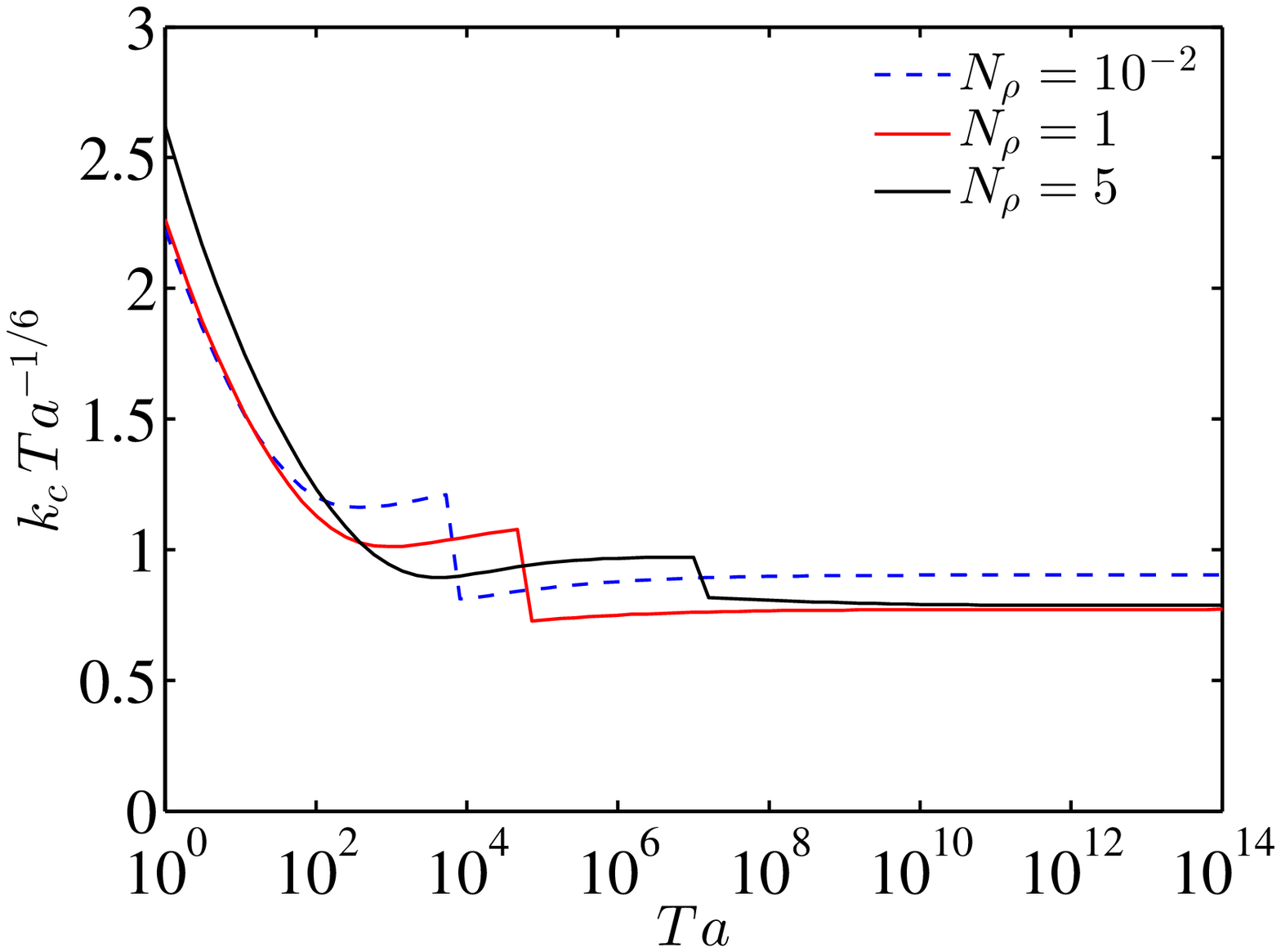}
      } \quad
      \subfigure[]{
      \includegraphics[width=7.5cm]{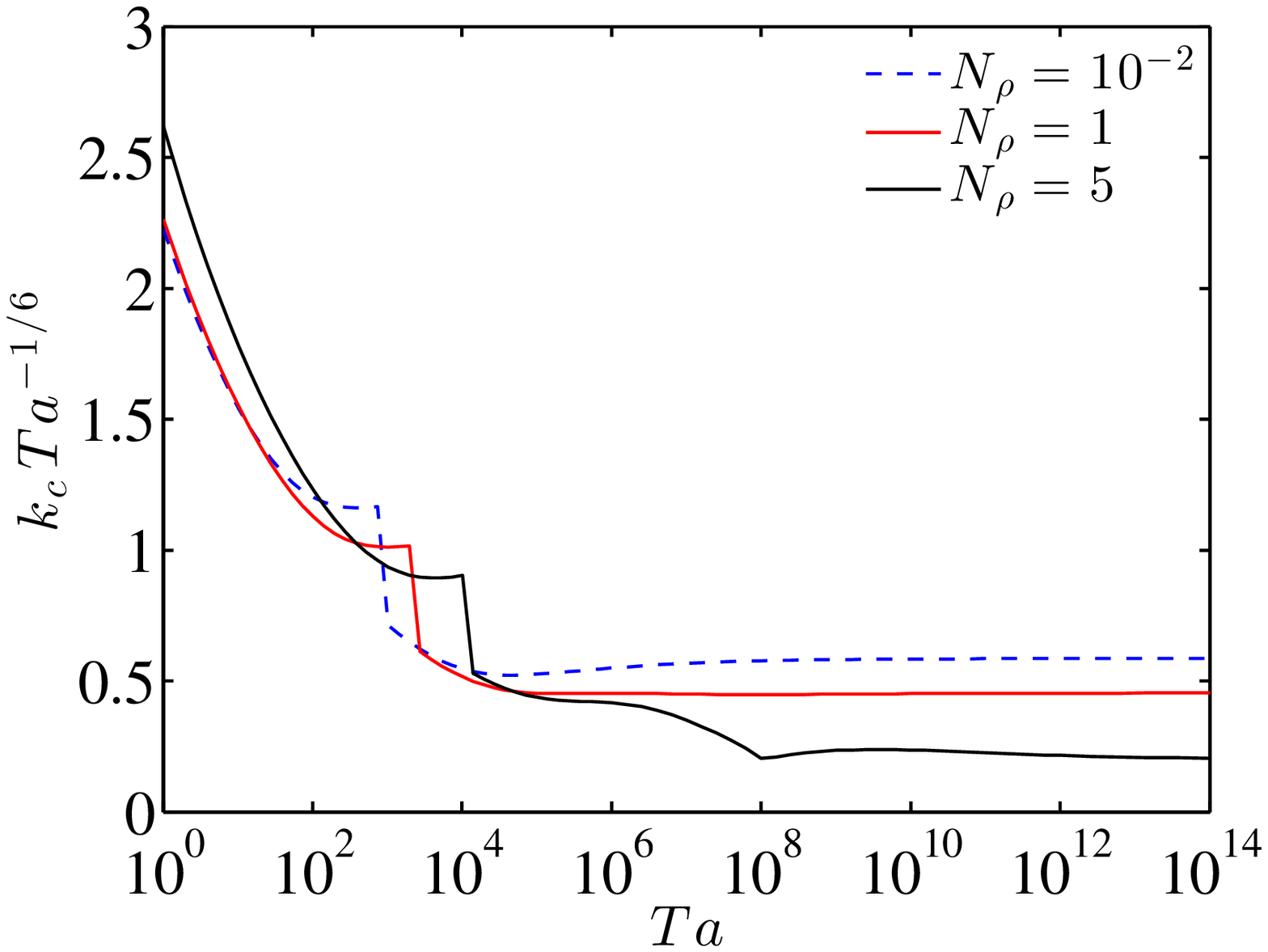}
      } \\   
     \subfigure[]{
      \includegraphics[width=7.5cm]{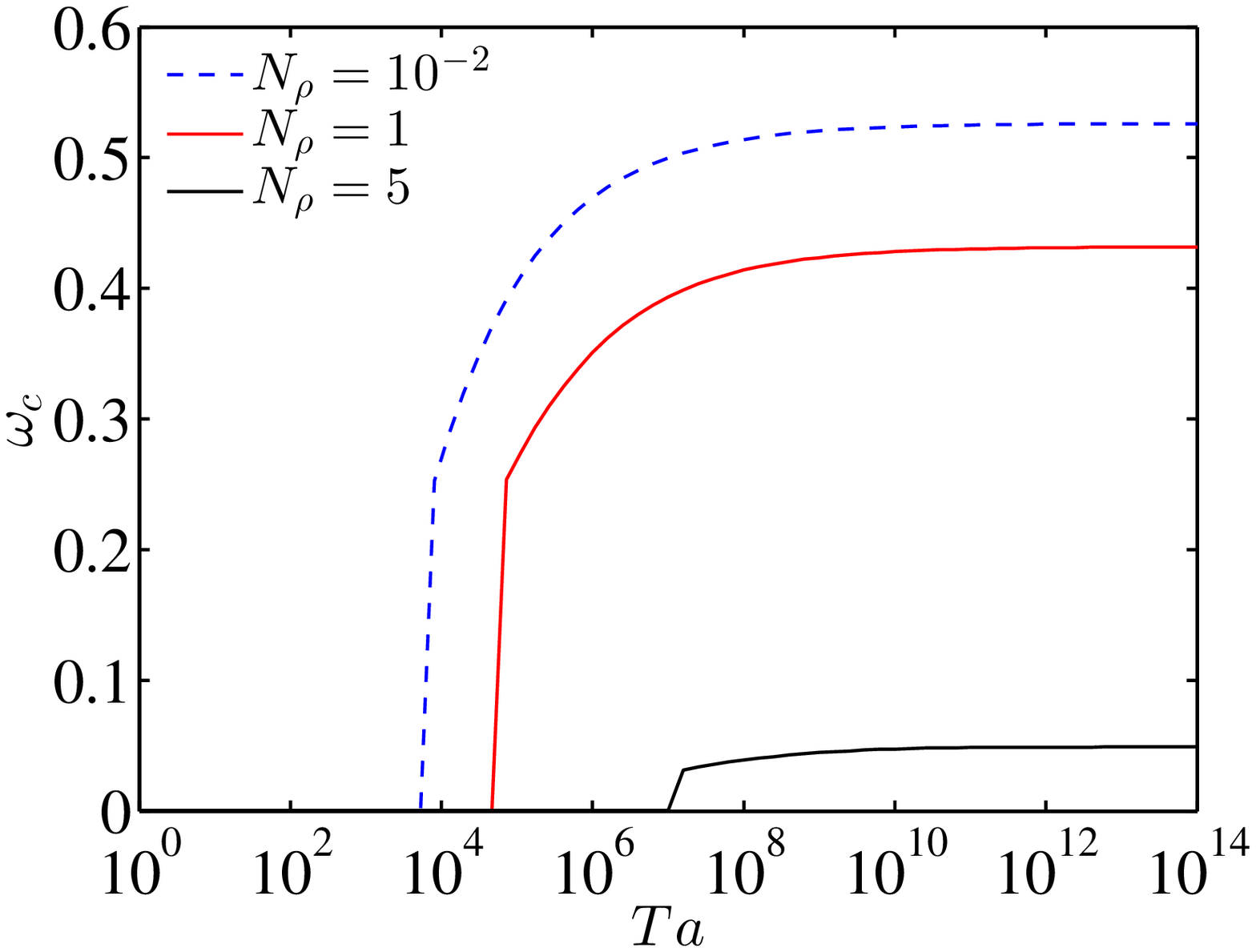}
      } \quad
      \subfigure[]{
      \includegraphics[width=7.5cm]{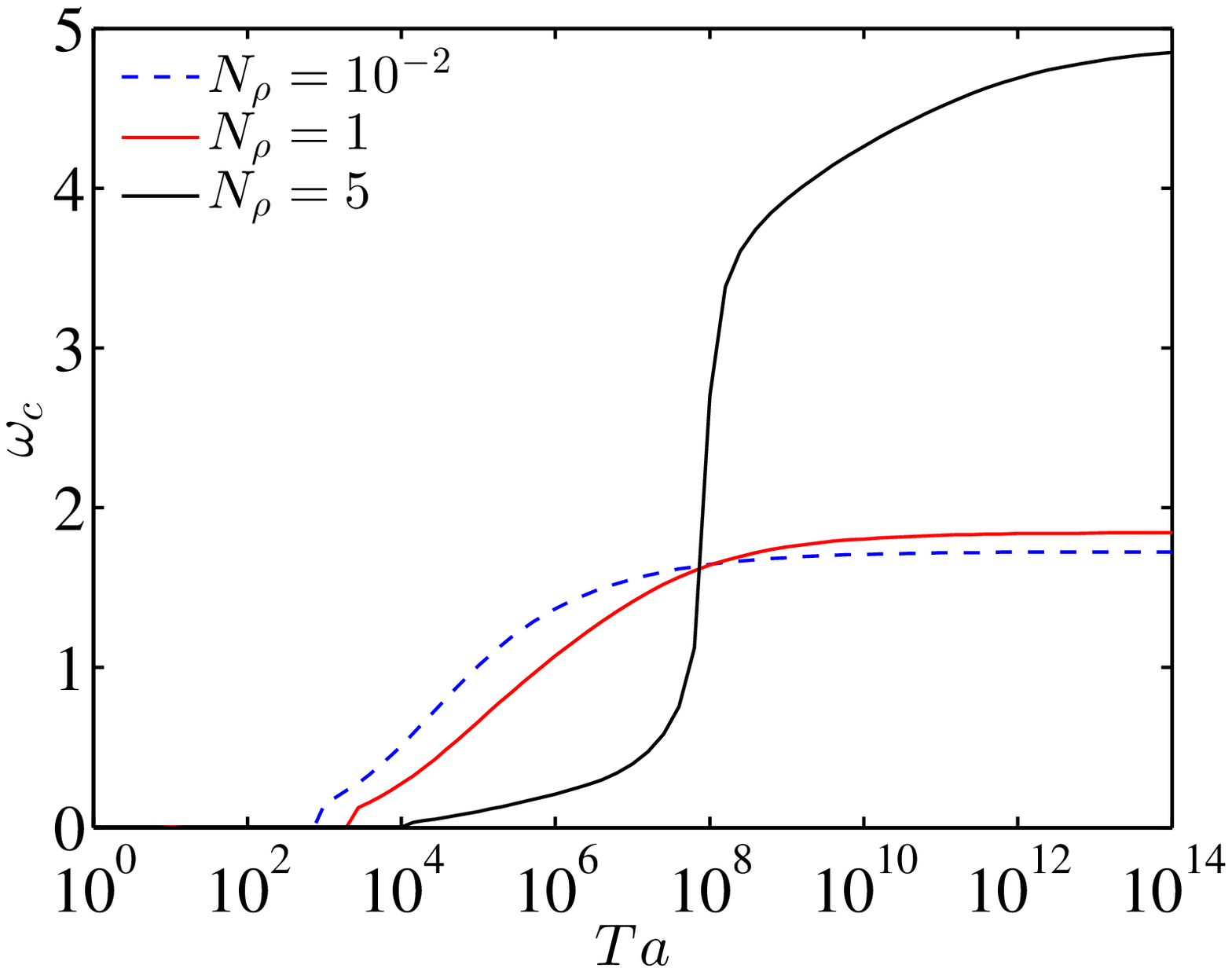}
      } \\   
  \end{center}
\caption{Critical parameters obtained from the Navier-Stokes equations (NSE) with polytropic index $n=1.49$, as a function of the Taylor number ($Ta$) for Prandtl numbers $Pr=0.5$ [(a), (c), (e)] and $Pr=0.1$ [(b), (d), (f)].   In each plot results are given for three different values of the background stratification parameter, $N_\rho$:  $N_\rho=10^{-2}$ (dashed blue curves); $N_\rho=1$ (solid red curves); $N_\rho=5$ (solid black curves).  The quantities shown are the critical Rayleigh number [(a),(b)], critical wavenumber [(c),(d)], and critical frequency [(e),(f)].  All values have been normalised by their respective asymptotic scalings valid in the limit $Ta \rightarrow \infty$.}
\label{F:CompCrit}
\end{figure}

Figure \ref{F:CompCrit} shows the critical parameters as a function of the Taylor number for two different Prandtl numbers and three different values of $N_\rho$ for the NSE; the case $N_\rho = 10^{-2}$ yields results close to those obtained from the OBE, whereas $N_\rho=5$ represents a relatively strong background stratification.  The first column in Figure \ref{F:CompCrit} [(a), (c), (e)] is for $Pr=0.5$ and the second column [(b), (d), (f)] is for $Pr=0.1$.    All the critical parameters are scaled by their respective large Taylor number asymptotic scalings \citep{sC61}.  Three different dynamical regimes can be distinguished in the critical parameters as the Taylor number is varied.  For sufficiently low Taylor number we observe a weak dependence of the critical parameters on the Taylor number and the convective motions are steady at onset; in this regime the Coriolis force is relatively weak with the dominant force balance occurring between pressure, viscosity and buoyancy forces.  Beyond this regime, for given values of $N_\rho$ and $Pr$, there is a finite value of $Ta$ at which a significant shift in critical parameter behavior is observed.  This effect is seen in the critical wavenumber curves [Figures \ref{F:CompCrit}(c) and (d)] as an abrupt shift towards lower critical wavenumber, and in the critical frequency curves [Figures \ref{F:CompCrit}(e) and (f)] as a shift to non-zero values of $\omega_c$.  The change in behavior is well known from the OBE and is associated with a change from steady convection to overstable oscillations \citep{sC61}.  It has been  shown that this regime occurs when $Pr \sim O(Ta^{-1/2})$ and is well described by a thermally-driven inertial wave in which the dominant horizontal force balance is between inertia, Coriolis, and pressure forces for $Ta$ sufficiently large \cite{kZ97b}.  As the Taylor number is increased further the rapidly rotating regime occurs in which the scaled critical parameters approach constant values and the dominant horizontal force balance is between pressure and Coriolis forces (i.e.~geostrophy).

\begin{figure}
  \begin{center}
     \subfigure[]{
      \includegraphics[width=7.5cm]{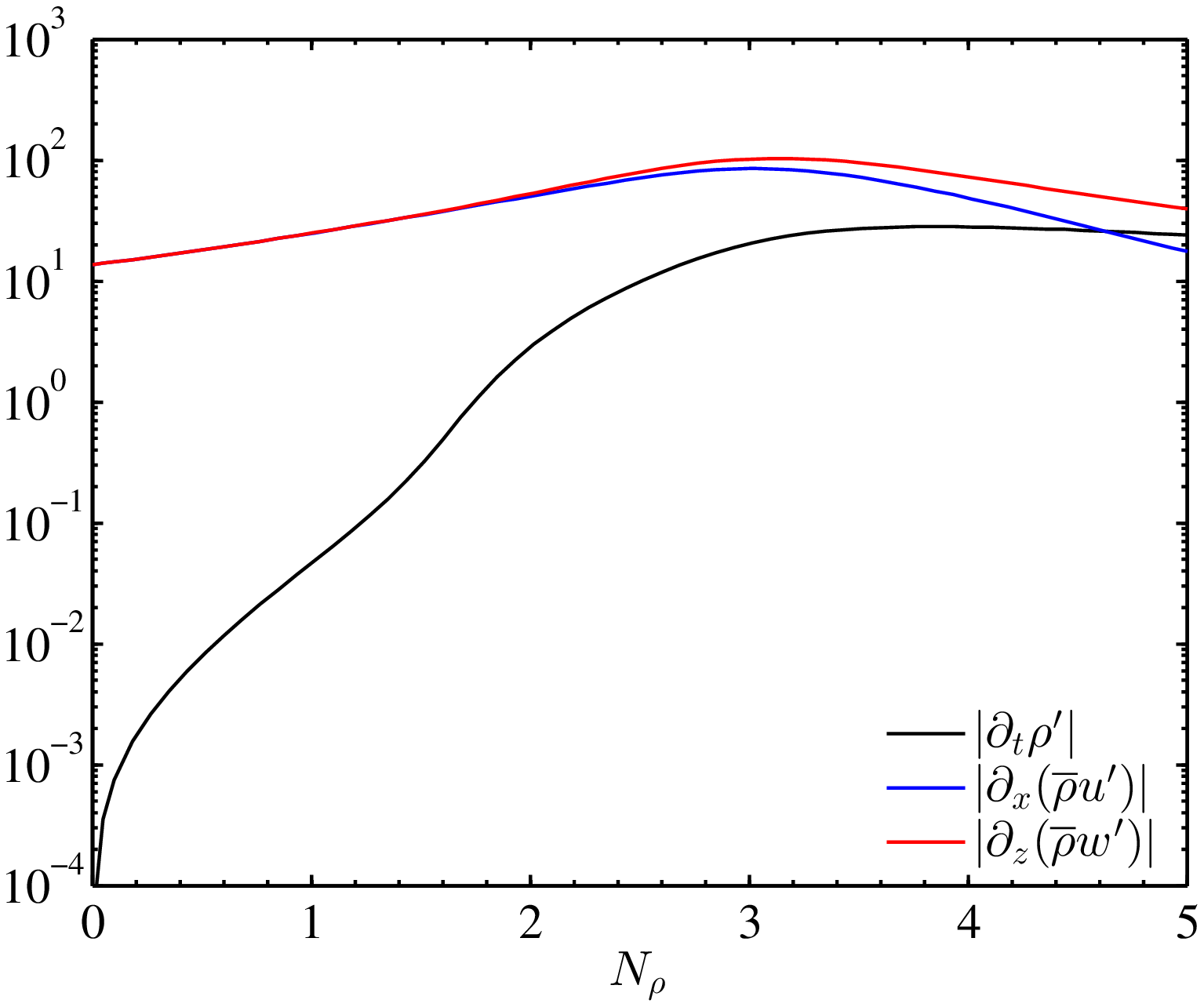}
      } \quad
      \subfigure[]{
      \includegraphics[width=7.5cm]{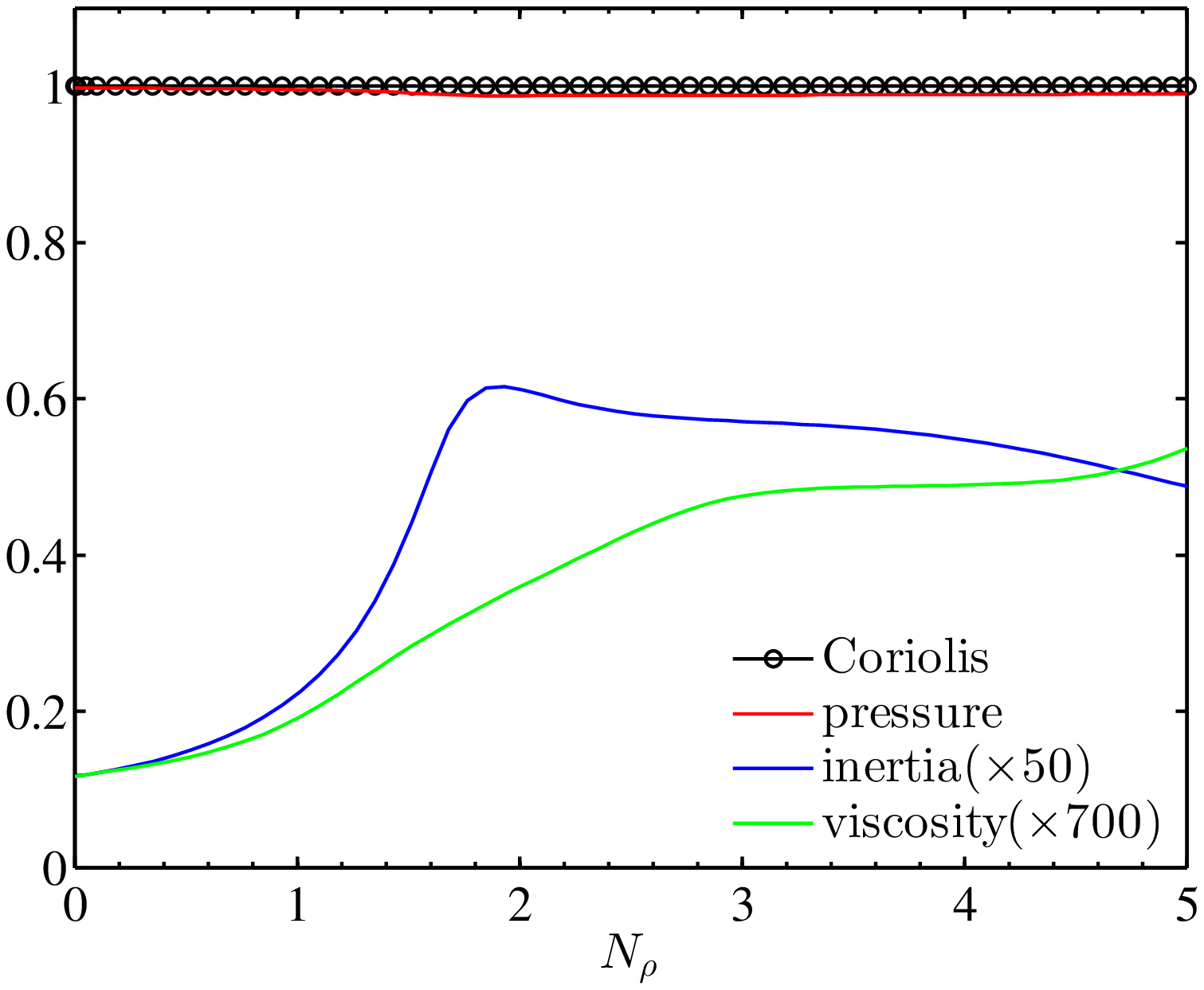}
      } \\   
  \end{center}
\caption{Magnitudes of the terms present in the (a) compressible conservation of mass equation and (b) the $x$-component of the NSE as a function of the background density stratification, $N_\rho$.  The axial norm of the terms in each equation is shown.  For simplicity convection rolls oriented in the direction of the $y$-axis are considered.  In both plots the Prandtl and Taylor numbers are fixed at $Pr=0.1$ and $Ta=10^{12}$, respectively.  Plot (a) shows the growth of the temporal derivative of the density perturbation as the background stratification is increased, whereas plot (b) shows that the eigenmodes remain geostrophically balanced (i.e., Coriolis and pressure balance) for all $N_\rho$ investigated.  Note that the magnitude of the inertial and viscous terms in (b) have been multiplied by large factors to show their behavior.}
\label{F:mass}
\end{figure}

The influence of stratification on the critical parameters is fundamentally dependent upon the Prandtl number.  For $Pr=0.5$, Figure \ref{F:CompCrit}(a) shows that $Ra_c$ increases with increasing $N_\rho$ for all Taylor numbers investigated.  For $Pr=0.1$, Figure \ref{F:CompCrit}(b) shows that the influence of stratification on $Ra_c$ is also dependent upon the Taylor number: for $Ta \lesssim 10^8$, stratification tends to stabilise the fluid layer and $Ra_c$ increases with increasing stratification; for $Ta \gtrsim 10^8$, $Ra_c$ decreases by nearly two orders of magnitude for $N_\rho=5$ in comparison to the $N_\rho=10^{-2}$ values.  The $Pr=0.5$ results are consistent with the $Pr$ order unity results given in \cite{mC14}, in the sense that stratification tends to have a stabilizing influence and leads to larger values of $Ra_c$.  The significant reduction of $Ra_c$ for $Pr=0.1$ shown in Figure \ref{F:CompCrit}(a) is also characterized by a discontinuity in the critical wavenumber curve at $Ta \approx 10^8$ given in Figure \ref{F:CompCrit}(d) and a significant increase in the critical frequency curve shown in Figure \ref{F:CompCrit}(f).  This behavior is associated with a rise in the magnitude of the temporal derivative of the density perturbation in the compressible mass conservation equation \eqref{E:mass}, despite the background state being nearly adiabatic; we discuss this point in more detail below [e.g.~see Figure \ref{F:mass}(a)].

The marginal curves given in Figures \ref{F:CompCrit}(e) and (f) for the critical frequency show that stratification tends to increase the Taylor number at which the instability becomes oscillatory (i.e., $\omega_c \ne 0$).  For $Pr=0.5$, we find that increasing stratification leads to a decrease in $\omega_c$ [Figure \ref{F:CompCrit}(e)], whereas for $Pr=0.1$ the opposite is true [Figure \ref{F:CompCrit}(f)].  In agreement with the OBE results \citep{sC61}, we find that decreasing the Prandtl number while holding all other parameters constant leads to an increase in the critical frequency. 

\begin{figure}
  \begin{center}
     \subfigure[]{
      \includegraphics[width=7.7cm]{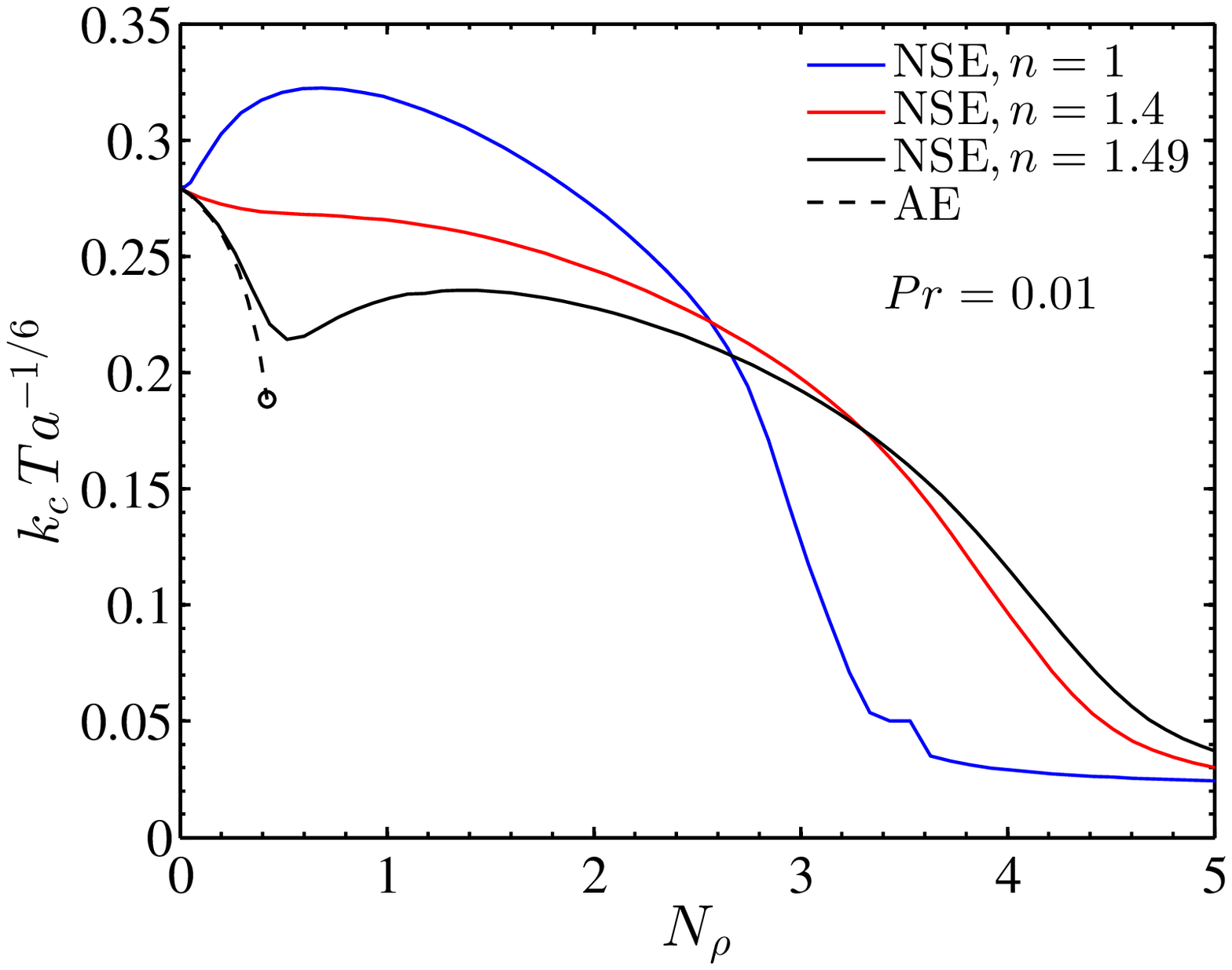}
      } \quad
      \subfigure[]{
      \includegraphics[width=7.5cm]{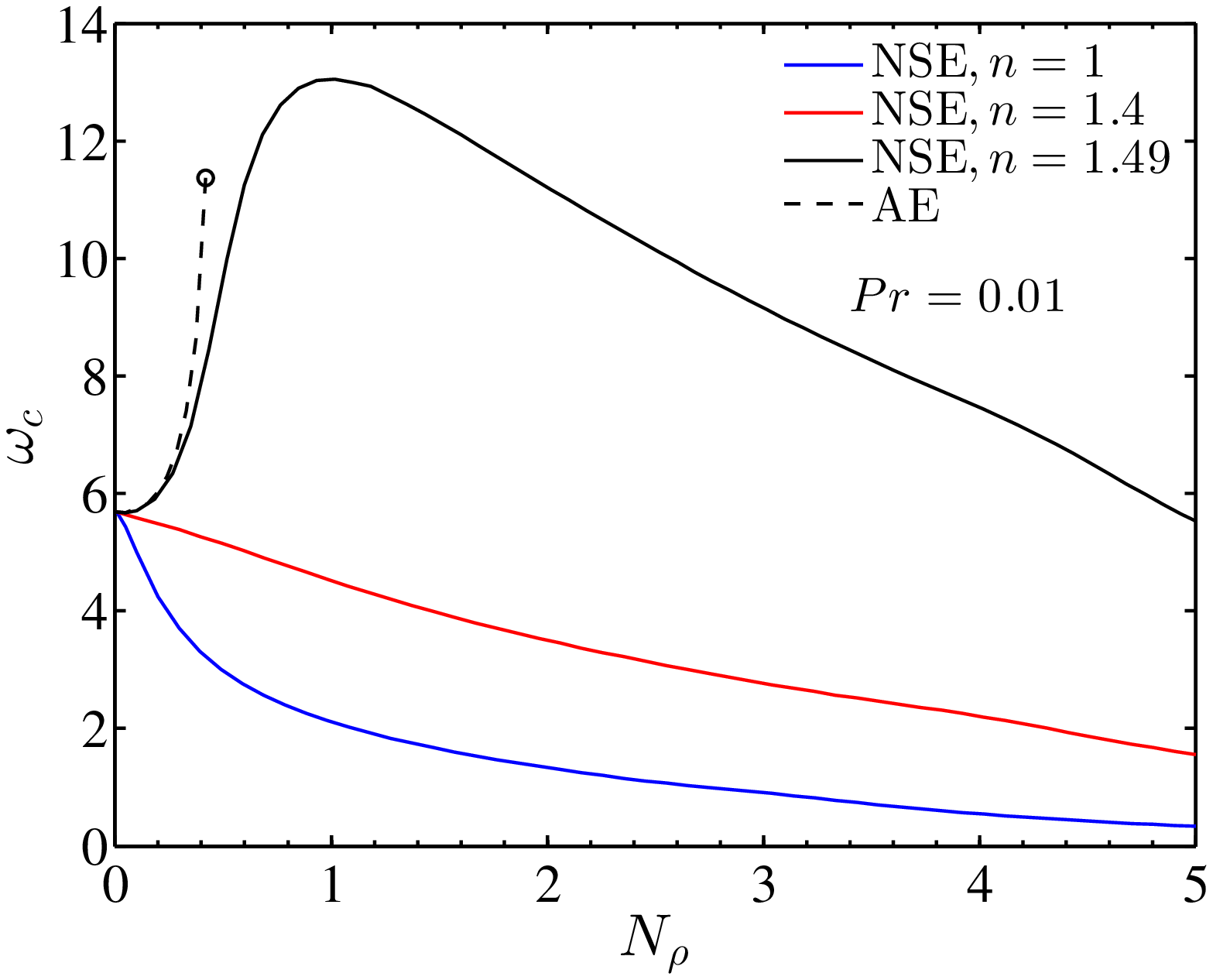}
      } \\   
    \subfigure[]{
      \includegraphics[width=7.7cm]{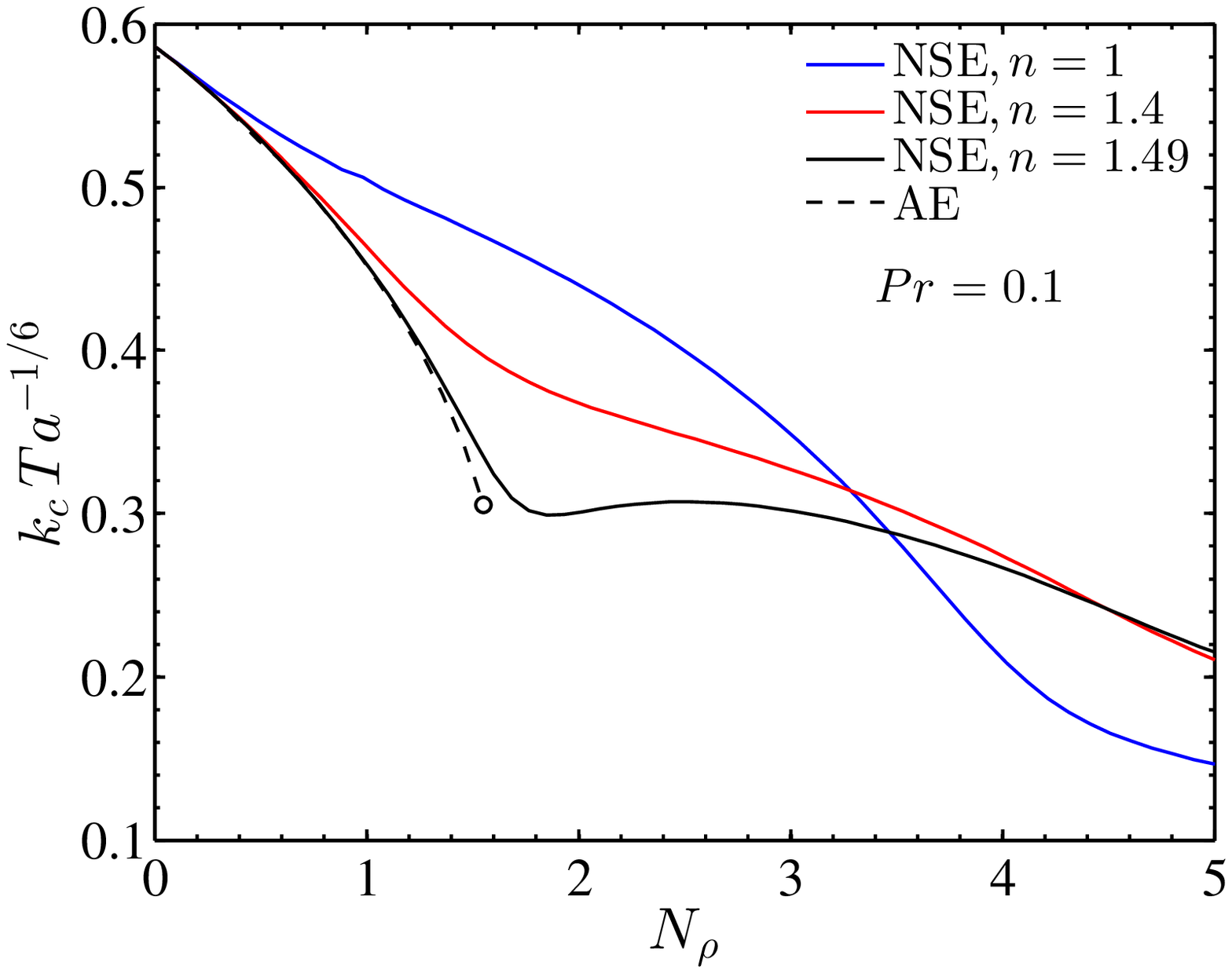}
      } \quad
      \subfigure[]{
      \includegraphics[width=7.5cm]{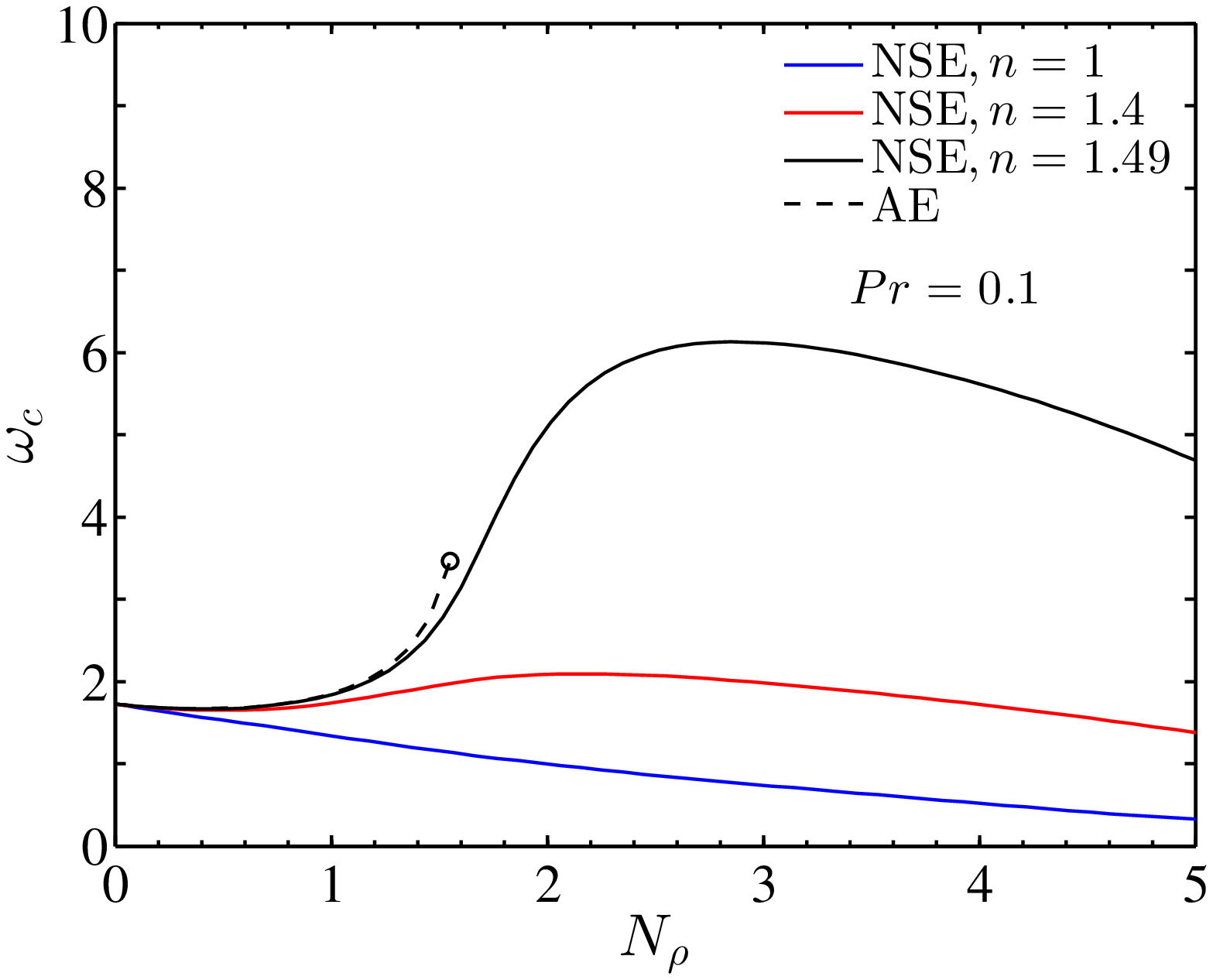}
      } \\   
     \subfigure[]{
      \includegraphics[width=7.5cm]{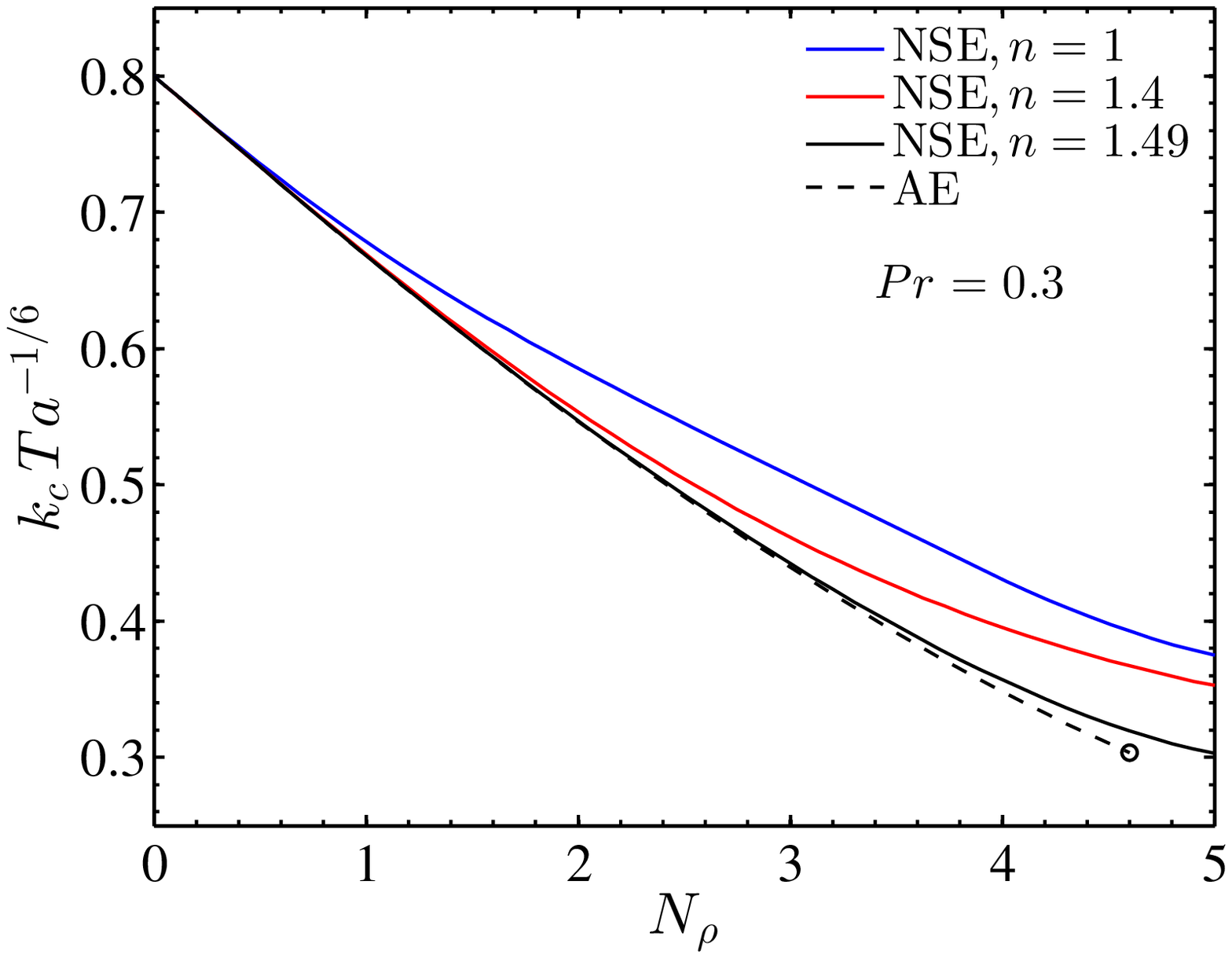}
      } \quad  
     \subfigure[]{
      \includegraphics[width=7.5cm]{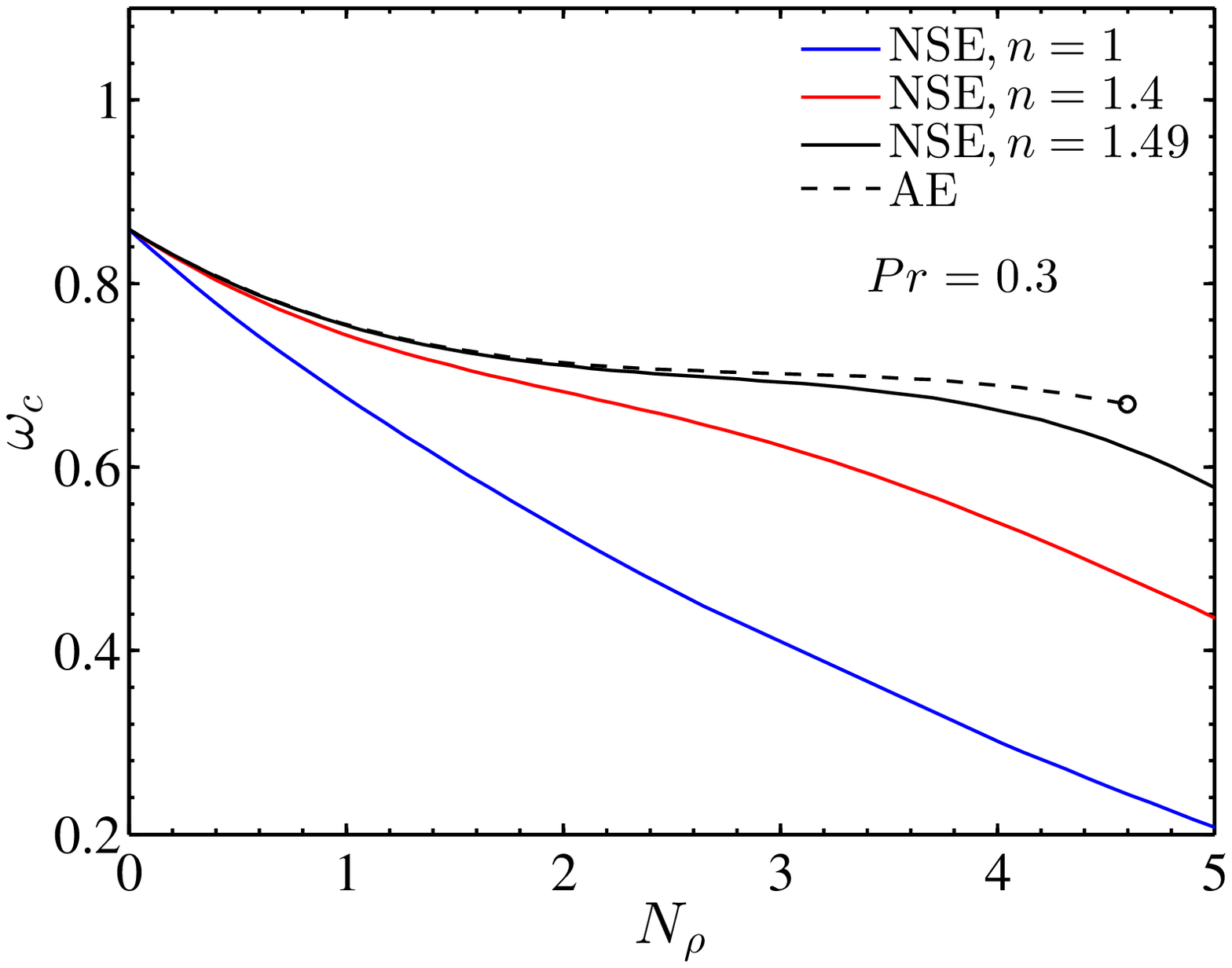}
      } \\        
  \end{center}
  \caption{Comparison of the critical wavenumber [(a),(c),(e)] and critical frequency [(b),(d),(f)] for the Navier-Stokes equations (NSE, solid curves) and anelastic equations (AE, dashed black curves) as a function of the background stratification $N_\rho$.  The first row [(a),(b)] are results for $Pr=0.01$, the second row [(c),(d)] are results for $Pr=0.1$ and the third row [(e),(f)] are results for $Pr=0.3$; for all cases the Taylor number is fixed at $Ta=10^{12}$.  The AE results diverge from the NSE results as the stratification level is increased, then fail at a finite value of $N_\rho$ (denoted by the black open circle).}
\label{F:CritNrho}
\end{figure}

In Figure \ref{F:CritNrho} we show the influence of stratification on the critical wavenumber and critical frequency for three values of $Pr$; results from both the NSE (solid curves) and the AE (dashed curves) are shown with the Taylor number fixed at $Ta=10^{12}$.  To illustrate the influence of deviating from an adiabatic background state, we show results for three different values of the polytropic index for the NSE that characterise the behavior of the critical parameters as $n \rightarrow n_a$.  In general, we find that the AE results diverge from the NSE results as $N_\rho$ increases.  For a given value of $Pr$, once $N_\rho$ was increased beyond a certain critical value, the AE yielded growing eigenmodes with negative Rayleigh numbers, in agreement with the work of \cite{sD95}.  Given that these modes do not exist in the NSE, we deem them unphysical and do not show them in any of the figures; the open black circles denote the final value of $N_\rho$ for which we found physically meaningful results (i.e.~$Ra_c > 0$).  For $Pr=0.01$ [Figures \ref{F:CritNrho}(a) and (b)], the AE produced critical values up to $N_\rho \approx 0.4$, and for $Pr=0.3$ the AE failed at $N_\rho \approx 4.6$.  The AE closely approximated the NSE $Pr=0.5$ results given in Figure \ref{F:CompCrit} for all $N_\rho$ investigated.  We thus find that the range of stratification for which the AE can approximate the NSE becomes increasingly small as the Prandtl number is reduced.  As first noted by \cite{sD95}, the Taylor number is required to be sufficiently large to observe the spurious behavior of the AE.  For $Pr=0.1$ and $N_\rho=5$, the AE yield critical parameters that closely approximate the NSE critical parameters up to $Ta \approx 10^8$.  For the large Taylor number limit, we find that the AE can accurately reproduce the NSE critical parameters for Prandtl numbers $Pr \gtrsim 0.5$.  Importantly, Figures \ref{F:CritNrho}(b), (d), and (e) show that $\omega_c$ is an increasing function of $n$ for low Prandtl numbers.

For Prandtl numbers $\lesssim 0.3$, the NSE results with $n=1.49$ exhibit inflection points in both the critical wavenumber curves [Figures \ref{F:CritNrho}(a) and (c)] and the critical frequency curves [Figures \ref{F:CritNrho}(b) and (d)].  As mentioned previously for Figure \ref{F:CompCrit}, we found that this behavior is due to the increase in magnitude of the temporal derivative of the density perturbation in the compressible mass conservation equation [see Figure \ref{F:mass}(a)].

Provided that the polytropic index $n$ is close to $n_a$, the differences in the background states of the two equation sets (e.g.~$\mrho$ and $\rho_0$) remain small and the observed failure of the AE for low Prandtl number gases can only be due to the different forms of the conservation of mass equation.  For simplicity, we will restrict our analysis to convection rolls that are oriented in the direction of the $y$-axis, such that $k_y \equiv 0$.  Equation \eqref{E:mass} then simplifies to 
\begineq
\dst \rho^{\prime} + \dsx \lb \mrho \pfu \rb + \dsz \lb \mrho \pfw \rb = 0 . \label{E:massx}
\endeq
In Figure \ref{F:mass}(a) we plot the axial norm of the magnitude of each of these three terms for the cases given in Figures \ref{F:CritNrho}(a) and (b) ($Pr=0.1$, $Ta=10^{12}$).  In agreement with the OBE, we find that $|\dst \rho^{\prime}| \rightarrow 0$ as $N_\rho \rightarrow 0$.  However, as $N_\rho$ is increased we observe that this term increases in magnitude until it becomes of comparable magnitude to the other terms present in the mass conservation equation. We find that the value of $N_\rho$ at which the AE fail occurs when $|\dst \rho^{\prime}| \approx O(0.1)$.  Some AE linear stability results were calculated with a $\dst \rho_1$ term included in the conservation of mass equation, and it was confirmed that positive critical Rayleigh numbers were obtained for these cases.  These results show that regardless of how close a compressible fluid system is to being adiabatically stratified, the AE will yield physically spurious results if the Prandtl number is much less than $Pr \approx 0.5$.

The asymptotic scalings evident in Figure \ref{F:CompCrit} suggests that the flow becomes geostrophically balanced as $Ta \rightarrow \infty$, even as the background stratification becomes large. In Figure \ref{F:mass}(b) we plot the axial norm of the magnitude of each term present in the $x$-component of the NSE,
\begineq
\begin{split}
\mrho \dst \pfu - Ro^{-1} \, \mrho \pfv = -  H_s \dsx \pp \, + \\ \sqrt{\frac{Pr}{Ra_o}} \lsq \nabla^2 \pfu + \frac{1}{3} \dsx \lb \dsx \pfu + \dsy \pfv  + \dsz \pfw \rb \rsq , \label{E:compmomx} 
\end{split}
\endeq
where we refer to each of the four terms present, beginning on the left hand side, as inertia, Coriolis, pressure, and viscosity, respectively.  For $y$-oriented convection rolls, the pressure gradient is identically zero in the $y$-component of the NSE and thus this equation is not considered.  The magnitude of each term was normalised by the magnitude of the Coriolis term at each value of $N_\rho$; the results show that geostrophy holds for all $N_\rho$ investigated with inertia and viscosity playing a subdominant role.  We emphasise that geostrophy is a \textit{point-wise} balance in the sense that it holds for all points in space; our use of axial norms is meant to simplify the analysis and it was confirmed that the balance indeed holds at all axial locations.    

\begin{figure}
  \begin{center}
     \subfigure[]{
      \includegraphics[width=7.6cm]{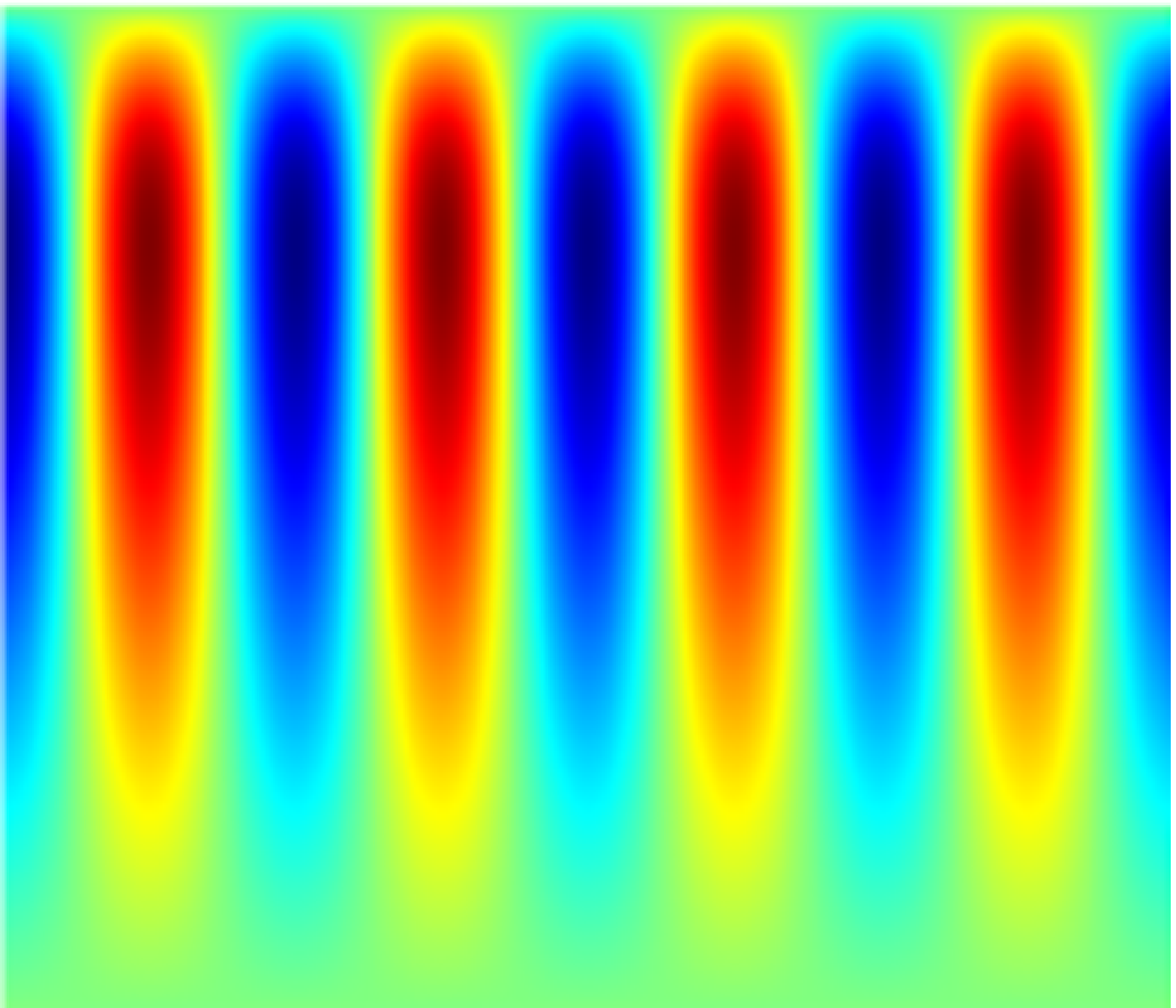}
      } \quad
      \subfigure[]{
      \includegraphics[width=7.6cm]{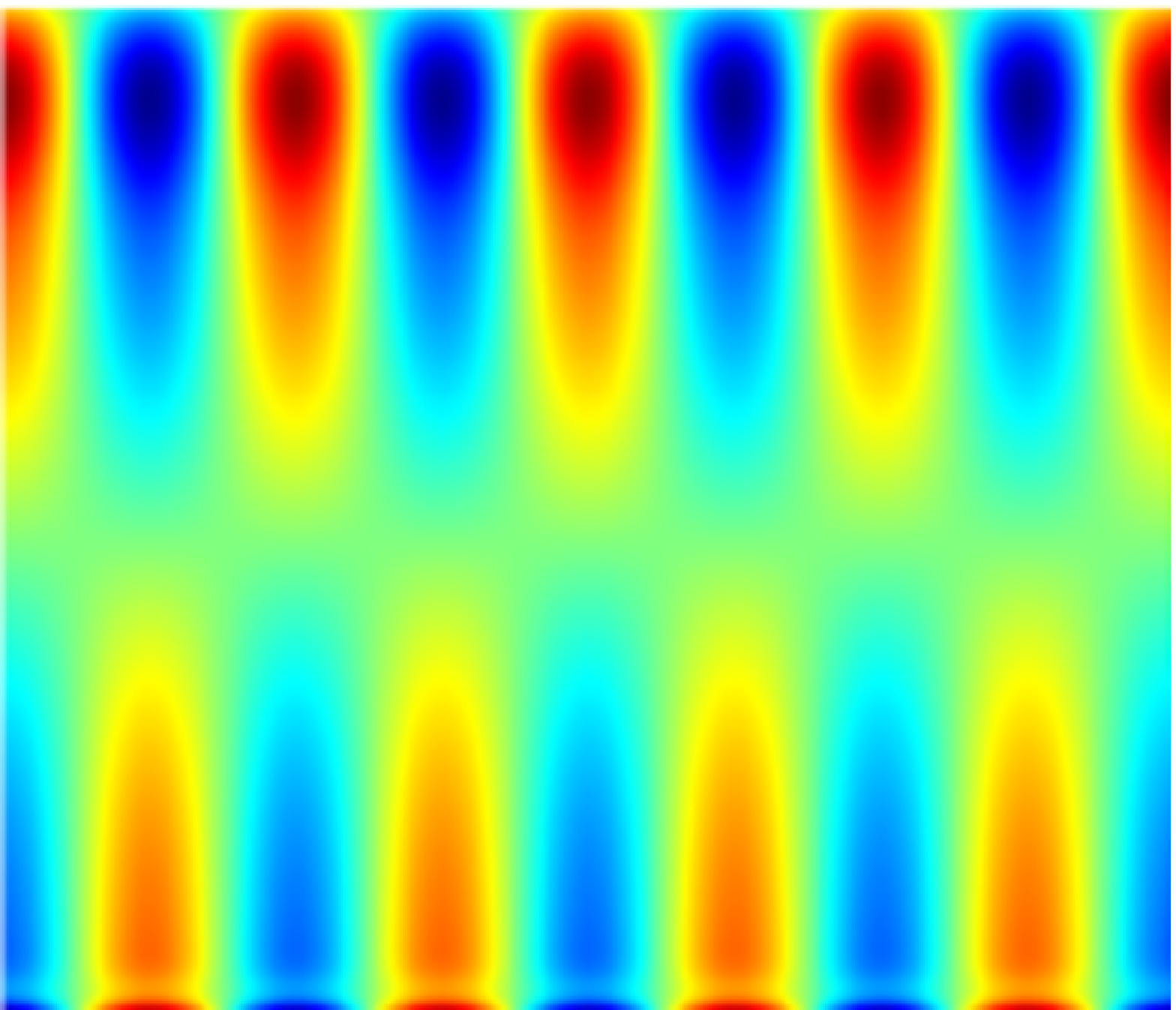}
      } \\   
  \end{center}
  \caption{Visualization of marginally stable compressible eigenmodes in the $x$-$z$ plane for $Pr=0.1$, $Ta=10^{12}$, $N_\rho=5$ and $n=1.49$. (a) Axial ($z$) velocity, $\pfw$; and (b) density perturbation, $\prho$.  The colour scale is of arbitrary magnitude with red representing positive values and blue representing negative values.}
\label{F:eigs}
\end{figure}

Visualizations of marginally stable compressible eigenmodes are given in Figure \ref{F:eigs} for $Pr=0.1$, $Ta=10^{12}$, $N_\rho=5$ and $n=1.49$.  We note that the AE fail for these parameter values and are thus incapable of reproducing these modes.  The horizontal dimension of each figure is scaled to include four unstable wavelengths.  In contrast to $Pr \sim O(1)$, Figure \ref{F:eigs}(a) shows that the axial velocity perturbation is shifted towards the top of the layer, with weaker axial motions near the base of the layer when the Prandtl number becomes small.  The density perturbation in Figure \ref{F:eigs}(b) also shows a strong peak near the top of the layer, but also possesses more complex, boundary-layer-type behavior near the base.  For these parameter values, both figures illustrate the highly anisotropic spatial structure indicative of high Taylor number convection \citep[e.g.][]{sC61}.  

\begin{figure}
  \begin{center}
     \subfigure[]{
      \includegraphics[width=7.6cm]{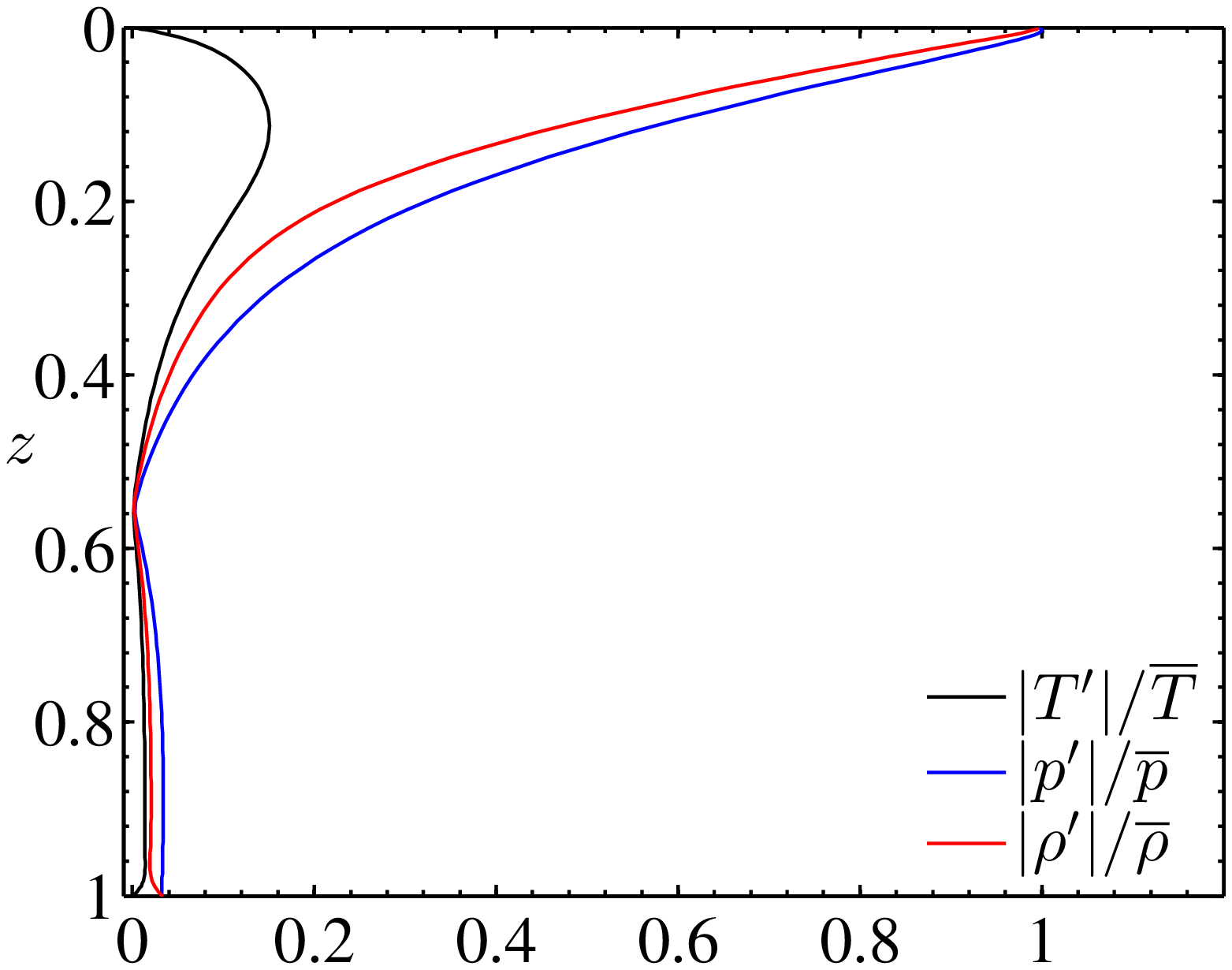}
      } \quad
      \subfigure[]{
      \includegraphics[width=7.6cm]{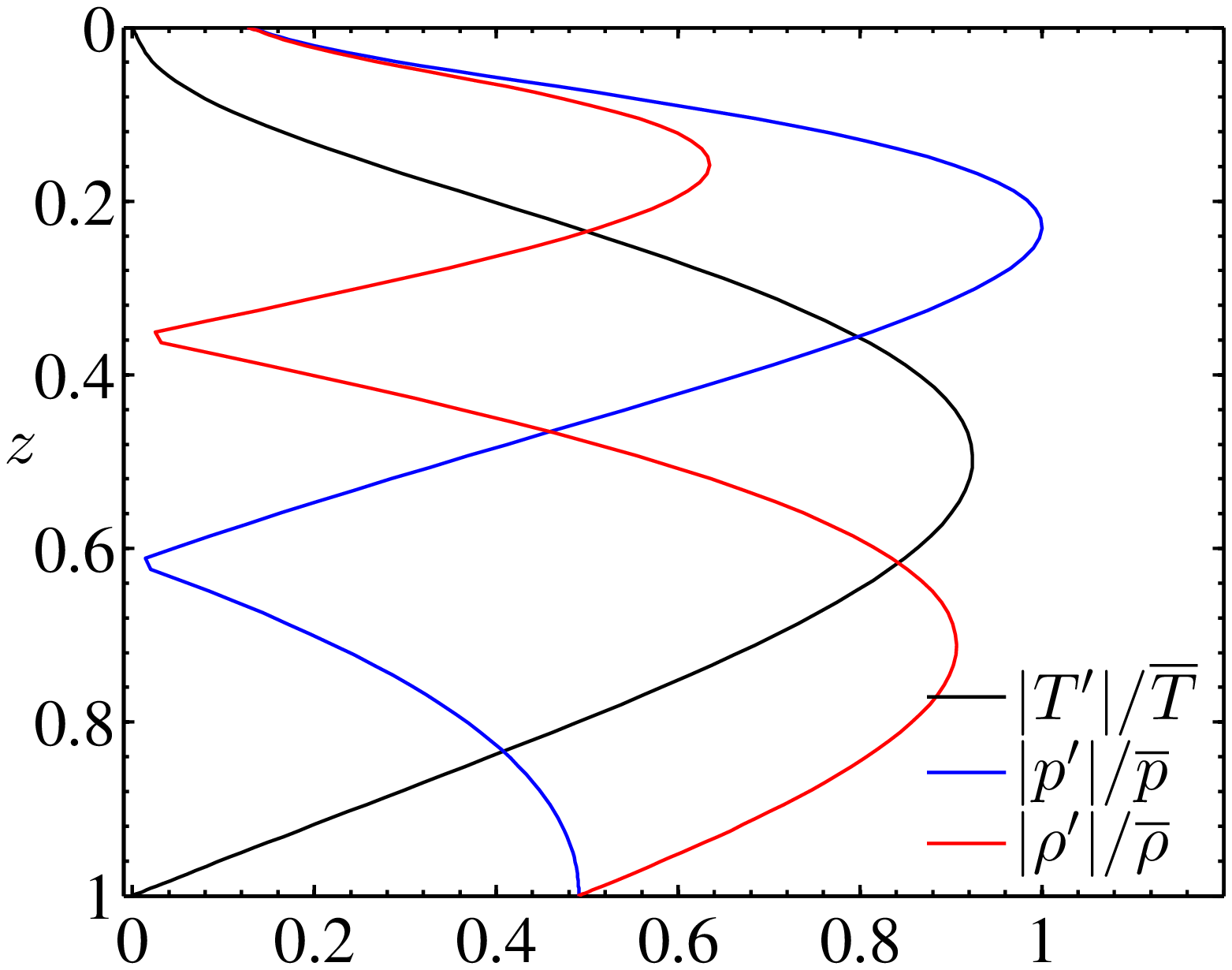}
      } \\   
  \end{center}
  \caption{Axial profiles of the magnitude of the three terms present in the linearised equation of state (perfect gas law) for (a) $Pr=0.1$ and (b) $Pr=1$.  Both cases are for $Ta=10^{12}$ and $N_\rho=5$.  Each thermodynamic perturbation variable has been scaled by the maximum value of the pressure perturbation for the two different cases shown.}
\label{F:gaslaw}
\end{figure}

A key distinction between incompressible convection, compressible convection in $Pr \gtrsim 1$ gases, and compressible convection in low Prandtl number gases is the spatial dependence and relative magnitudes of the state variables.  For linear motions, the gas law becomes
\begineq
\frac{\pp}{\mp} = \frac{\prho}{\mrho} + \frac{\pT}{\mT}.
\endeq
In Figure \ref{F:gaslaw} we present axial profiles of the magnitude of each term appearing in the above equation for $Ta=10^{12}$, $N_\rho=5$ and $n=1.49$.  For comparison, we show results for both $Pr=0.1$ [Figure \ref{F:gaslaw}(a)] and $Pr=1$ [Figure \ref{F:gaslaw}(b)] .  Note that because our analysis is restricted to linear convection, the amplitude of the convective perturbations is arbitrary and each variable has been scaled by the maximum value of the pressure perturbation.  Both plots show distinct spatial behavior with the (normalised) pressure and density perturbations of comparable magnitude, and largest in the upper portion of the layer for $Pr=0.1$.  The temperature perturbations for the $Pr=0.1$ case are an order of magnitude smaller than the pressure and density in the upper portion of the layer, but all three terms are of comparable magnitude in the lower portion of the layer.  Figure \ref{F:gaslaw}(b) shows that for $Pr=1$ all three terms are of comparable magnitude throughout the depth of the layer.  These results should be contrasted with incompressible convection characterised by $(H_s^{-1}, H_a^{-1}) \rightarrow 0$; in this case $\pp/\mp \rightarrow 0$, $(\mrho, \mT) \rightarrow 1$, and thus $\prho \rightarrow -\pT$.  Pressure perturbations therefore play no role in the buoyancy force in the incompressible limit. 

\section{Discussion}

An interesting property of rapidly rotating compressible convection is that it implicitly resides in the low Mach number regime, despite the characteristics of the background state.  Recall that the Mach number is defined as the ratio of the characteristic flow velocity to the speed of sound
\begineq
Ma = \frac{U}{c},
\endeq
where for a perfect gas the dimensional speed of sound at the top boundary is given by $c = \sqrt{\gamma p_o/\rho_o}$ \citep[e.g.][]{dL13}.  In the limit of rapid rotation, the fluid velocity can be scaled with the viscous diffusion acting over the small horizontal length scale $L$ such that $U \sim \mu / (\rho_o L)$ \citep[e.g.][]{sC61,mS06}. The Mach number can then be written as
\begineq
Ma = \lb \frac{H}{L} \rb \lb \frac{n_a Pr}{H_s H_a Ra_o} \rb^{1/2} ,
\endeq
where the factor $H/L$ is simply the critical wavenumber. Assuming the strength of the background stratification to be order one in magnitude, such that the prefactor $(n_a / H_a H_s) \sim O(1)$, two different bounds on the Mach number can then be derived depending upon the magnitude of the Prandtl number.  For $Pr \sim O(1)$ we have
\begineq
Ma \sim Ta^{-1/6}, \quad (Pr \sim O(1)) 
\endeq
and the corresponding result for small Prandtl number is
\begineq
Ma \sim Pr^{1/6} Ta^{-1/6}. \quad (Pr \ll 1)
\endeq
In both of these relations we have used the fact that $k_c \sim Ta^{1/6}$ and $Ra_c \sim Ta^{2/3}$ for order one Prandtl numbers and for $Pr \ll 1$ we have $k_c \sim Pr^{1/3} Ta^{1/6}$ and $Ra_c \sim Pr^{4/3} Ta^{2/3}$. The above relations show that for $Ta \gg 1$ we always have $Ma \ll 1$.

Some important conclusions can be drawn from the balances shown in Figure \ref{F:mass}.  We first note that geostrophy is not restricted to the case of convection rolls oriented in the direction of the $y$-axis , and in the following we thus generalise to $k_y \ne 0$. Figure \ref{F:mass}(b) shows that the primary force balance in the horizontal dimensions is geostrophy
\begineq
Ro^{-1} \hz \times \mrho \ubp_g \approx - H_s \nabla_\perp \pp , \label{E:geo} \quad
\endeq
where $\nabla_\perp = \lb \dsx, \dsy, 0 \rb$ and the geostrophic velocity components are denoted by $(\pfu_g, \pfv_g)$.  Taking the curl of equation \eqref{E:geo} shows that we have horizontally non-divergent flow at leading order,
\begineq
\divp \lb \mrho \ubp_g \rb \approx 0 \quad \Rightarrow \quad \divp \ubp_g \approx 0 . \label{E:div}
\endeq
However, convection requires that the geostrophic balance is perturbed in the sense that small $O(Ro)$ ageostrophic motions must be present.  This implies that mass can only be conserved if we consider the $O(Ro)$ correction to the mass conservation equation, i.e.
\begineq
\dst \prho + \divp \lb \mrho \ubp_{ag} \rb + \dsz \lb \mrho \pfw \rb = 0, \label{E:ageo}
\endeq
where the ageostrophic $O(Ro)$ velocity field is denoted by $\ubp_{ag}$.  In fact, the terms for which the magnitudes are given in Figure \ref{F:mass}(a) are precisely those present in equation \eqref{E:ageo} since \eqref{E:div} is trivially satisfied for geostrophically balanced convection.  Thus, equations \eqref{E:massx} and \eqref{E:ageo} are equivalent.  

The vortex stretching mechanism represents a fundamental component of quasi-geostrophic flows, and is the sole result of mass conservation associated with the ageostrophic velocity \citep{jP87,kJ06}.  For incompressible rapidly rotating convection, both in the plane layer geometry \citep{kJ06} and spherical geometries \citep{cJ00,eD04}, equation \eqref{E:ageo} simplifies to
\begineq
\divp \ubp_{ag} = - \dsz \pfw . \label{E:stretch_inc}
\endeq
For compressible rapidly rotating convection, equation \eqref{E:ageo} shows that vortex stretching is now due to the horizontal divergence of the ageostrophic momenta such that
\begineq
\divp \lb \mrho \ubp_{ag} \rb = - \dsz \lb \mrho \pfw \rb - \dst \prho . \label{E:stretch_c}
\endeq
Our results thus show that fluid compression is now an important  source (or sink) of axial vorticity that cannot be neglected in the governing equations, resulting in the presence of both longitudinal (compressional) waves and transverse, low frequency inertial waves.  

A non-dimensional scale analysis of equation \eqref{E:ageo} shows that for the three terms to be of comparable magnitude we must have
\begineq
|\omega_c \prho | \sim |k_c \ubp_{ag}| \sim | \pfw| , 
\endeq
since $|\mrho| \sim O(1)$ and $\dsz \sim O(1)$.  Furthermore, on noting $Ra_c \sim Ta^{2/3}$, our results show that $k_c \sim Ta^{1/6} \sim Ro^{-1}$, $|\ubp_{ag}| \sim O(Ro)$, and $|\pfw| \sim O(1)$.  We thus have 
\begineq
|\omega_c \prho|  \sim O(1). 
\endeq
Figure \ref{F:gaslaw} shows that $\prho \sim \pp$; from the geostrophic balance we can estimate $\pp$ as 
\begineq
\pp \sim \frac{1}{H_s Ro k_c} \sim \frac{1}{H_s} \quad \Rightarrow \quad \prho \sim \frac{1}{H_s} 
\endeq
We then have 
\begineq
\left|\frac{\omega_c}{H_s}\right|  \sim O(1),
\endeq
where we note that no assumptions have been made with regard to the size of $H_s$.  It is instructive to relate $H_s$ to the polytropic index.  We can define the deviation of the polytropic index from the adiabatic value by 
\begineq
\ep_n = \frac{n_a - n}{n_a} .
\endeq
For $\ep_n \ll 1$ we can then write 
\begineq
\frac{1}{H_s} \approx \frac{\ep_n}{\gamma H_a} ,
\endeq
such that 
\begineq
\left| \frac{\ep_n \omega_c}{\gamma H_a} \right| \sim O(1). \label{E:freq_scale}
\endeq
In order for the AE to be a reasonable approximation of the NSE, we must therefore have 
\begineq
\left| \frac{\ep_n \omega_c}{\gamma H_a} \right| \ll 1. \label{E:freq_AE}
\endeq
In general, the critical frequency $\omega_c$ is a function of all the dimensionless parameters ($Pr$, $N_\rho$, $n$, etc.).  Within our range of investigated parameters, however, we find that $\omega_c \sim O(0.1-1)$.  Taking the (non-dimensional) critical frequency $\omega_c \sim O(1)$ for simplicity we can now see from \eqref{E:freq_AE} that as observed the AE becomes a worse approximation as $N_\rho$, and thus $H_a^{-1}$ increases, despite having $\ep_n \ll 1$.  The fact that we have observed complete failure of the AE shows that no matter how small we take $\ep_n$ the AE will fail for finite values of $N_\rho$.  We find that this failure results from the fact that $\omega_c$ is an increasing function of the polytropic index.  Although it was previously assumed that $\partial_t \prho$ decreases as $H_s \rightarrow \infty$, our results show that this is not the case when the Prandtl number is small; any increase of $n$ in the NSE is associated with a concomitant increase in $\omega_c$.  Furthermore, $\omega_c$ also increases with decreasing $Pr$, and relation \eqref{E:freq_AE} shows that the range of $N_\rho$ over which the AE will provide accurate results approaches zero.

\section{Conclusions}

While the AE have been employed for numerous investigations of both gravity wave and convection dynamics, the results of the present work show that there exist severe limitations on their accuracy for high Taylor number and low Prandtl number fluids.  Prandtl numbers for typical gases range from $Pr\approx0.1$--$1$, plasmas have $Pr\approx10^{-6}$--$10^{-3}$, with liquid metals $Pr\approx10^{-3}$--$10^{-1}$.  The Taylor numbers for all natural systems are enormous, with the Earth's liquid core having $Ta \approx 10^{30}$ and the Sun's convection zone of the same order \citep{mM05}.  We find that the temporal derivative of the density perturbation is an elementary component of such systems, despite the smallness of the Mach number and the quasi-adiabaticity of the background state.  The resulting convective instabilities are characterised by compressional quasi-geostrophic oscillations and in this respect show that no strictly \textit{soundproof} model will be accurate for rapidly rotating compressible convection. Although the present work has for simplicity focused only on calorically perfect gases, we see no reason why our results will not be relevant for low Prandtl number liquids \citep[][]{aA05}.  It is therefore unlikely that the AE will be adequate for accurately modeling many systems; compressible simulations \citep[e.g.][]{nH02} and the development of new reduced equations will be necessary.  

It has long been known that the linear properties of rotating incompressible convection in low Prandtl number fluids shows distinct dynamics in comparison to $Pr \sim O(1)$ fluids.  Experimental \citep{eK13} and numerical \citep{mC12b} investigations show that these differences also extend into the nonlinear regime.  Our investigation has shown that the Prandtl number is also an important parameter in compressible convection, and has shown the inadequacies of the AE. Furthermore, recent numerical simulations also show that the Prandtl number has a controlling influence on dynamo action and may impact the morphology of both planetary and stellar magnetic fields \citep{cJ14}.  

Our findings have raised some important new questions concerning the validity of the AE that can only be addressed with future nonlinear investigations.  For instance, do the AE hold for $Pr \sim O(1)$ fluids when the flow field is turbulent and characterised by a broadband frequency and wavenumber spectrum?  Do the AE accurately model \textit{non-rotating} turbulent convection?  Answering these and other questions requires one-to-one comparisons between nonlinear simulations of both the AE and NSE and will be helpful for the development of more accurate convection models for geophysical and astrophysical fluids.

\begin{table}[ht]
  \caption{Comparison of linear stability data for the Oberbeck-Boussinesq equations (OBE), compressible Navier-Stokes equations (NSE) and the anelastic equations (AE).  The critical Rayleigh number, critical wavenumber, and critical frequency are denoted by $Ra_c$, $k_c$, and $\omega_c$, respectively.  $n$ is the polytropic index, $N_\rho$ is the number of density scale heights, and $Pr$ is the Prandtl number.  For all values listed the Taylor number is fixed at $Ta=10^{12}$.  All results with polytropic indices $n < 1.5$ were obtained from the NSE, whereas values listed with $n=1.5$ were obtained from the AE.  Values listed with $n=$``$-$" and $N_\rho=0$ correspond to results obtained from the Oberbeck-Boussinesq equations \citep[e.g.~see][]{sC61}.  AE results in which the critical parameters are not listed indicates a parameter range for which the AE failed to produce physically meaningful results.}
\begin{center}
\begin{tabular}{lcccccc}
$n$ & $N_\rho$ & $Pr$ & $Ra_c$ & $k_c$ & $\pm \omega_c$ \\
\hline
$-$ & 0 & 0.5 & $6.04\times10^8$ & 90.44 & $5.27\times10^{-1}$  \\
$-$ & 0 & 0.3 & $3.21\times10^8$ & 80.00 & $8.59\times10^{-1}$  \\
$-$ & 0 & 0.1 & $7.84\times10^7$ & 58.63 & $1.72$  \\
$1$ & 5 & 0.5 & $6.61\times10^8$ & 60.67 & $8.62\times10^{-2}$  \\
$1.4$ & 5 & 0.5 & $1.10\times10^9$ & 67.74 & $8.27\times10^{-2}$  \\
$1.49$ & 5 & 0.5 & $1.17\times10^9$ & 78.65 & $4.88\times10^{-2}$  \\
$1.5$ & 5 & 0.5 & $1.18\times10^9$ & 78.82 & $4.84\times10^{-2}$  \\
$1$ & 5 & 0.3 & $2.16\times10^8$ & 37.51 & $2.08\times10^{-1}$  \\
$1.4$ & 5 & 0.3 & $4.14\times10^8$ & 35.26 & $4.36\times10^{-1}$  \\
$1.49$ & 5 & 0.3 & $4.56\times10^8$ & 30.28 & $5.77\times10^{-1}$  \\
$1.5$ & 5 & 0.3 & $-$ & $-$ & $-$  \\
$1$ & 5 & 0.1 & $7.95\times10^6$ & 14.65 & $3.22\times10^{-1}$  \\
$1.4$ & 5 & 0.1 & $1.11\times10^7$ & 21.05 & $1.37$  \\
$1.49$ & 5 & 0.1 & $1.55\times10^6$ & 21.50 & $4.69$  \\
$1.5$ & 5 & 0.1 & $-$ & $-$ & $-$  \\
\end{tabular}
\end{center}
\label{T:stability}
\end{table}%

\section*{Acknowledgments}
This work was supported by the National Science Foundation under grants EAR \#1320991 (MAC and KJ) and EAR CSEDI \#1067944 (KJ and PM).

\bibliography{CompressBib}

\end{document}